\documentclass[12pt]{article}  
\usepackage{a41}
\usepackage{cite}
\usepackage{amsmath}
\usepackage{amssymb}
\usepackage{color}  

\usepackage[rflt]{floatflt}              
\usepackage{float}
\usepackage{slashed}

\def\b0{\beta_0}

\usepackage{array}

\newcommand{\spp}{s''}
\renewcommand{\sp}{s^\prime}

\newcommand{\HA}{{\rm H}}

\usepackage[english]{babel}
\usepackage[latin1]{inputenc}
\usepackage[T1]{fontenc}
\usepackage{ae}

\usepackage{url}

\usepackage{amsmath, amsthm, amssymb}
\newtheorem{thm}{Theorem}[section]

\newtheorem{definition}[thm]{Definition}

\newcommand{\Li}{{\rm Li}}
\newcommand{\HT}{{\rm H}^*}

\newcommand{\ep}{\varepsilon}

\usepackage{rotating}

\usepackage{graphicx}

\newcounter{mmacnt}
\def\restartmma{\setcounter{mmacnt}{0}}
\restartmma \catcode`|=\active
\def|#1|{\mathrm{#1}}
\catcode`|=12
\newenvironment{mma}{
 \par\smallskip
 \catcode`|=\active
 \parskip=0pt\parindent=0pt 
 \small
 \def\In##1\\{%
\def\linebreak{\hfill\break\null\qquad}%
\refstepcounter{mmacnt}
\hangindent=2.5em\hangafter=0
\leavevmode
\llap{\tiny\sffamily n[\arabic{mmacnt}]:=\kern.5em}%
\mathversion{bold}\footnotesize$\displaystyle##1$\normalsize
\mathversion{normal}\par
 }%
 \def\Print##1\\{%
\def\linebreak{\hfill\break}%
\hangindent=2.5em\hangafter=0
\leavevmode ##1\par}%
 \def\Out##1\\{%
\def\linebreak{$\hfill\break\null\hfill$}%
\kern\abovedisplayskip\par
\hangindent=2.5em\hangafter=0
\leavevmode
\llap{\tiny\sffamily Out[\arabic{mmacnt}]=\kern.5em}
\footnotesize$\displaystyle##1$\normalsize\hfill\null\par
\kern\belowdisplayskip
 }%
 \def\Warning##1##2\\{%
\def\linebreak{\hfill\break}%
\hangindent=2.5em\hangafter=0
\leavevmode
{\scriptsize##1 : ##2}\par}%
}{%
 \par\smallskip
}

\usepackage{color}

\newenvironment{fshaded}{%
\MakeFramed {\FrameRestore}
}%
{\endMakeFramed}

\def\b0{\beta_0}

\def\Gp0{{\Gamma^{'}_0}}
\def\Gp1{{\Gamma^{'}_1}}
\def\Gp2{{\Gamma^{'}_2}}

\allowdisplaybreaks[4]

\begin{document}
\setlength{\baselineskip}{0.515cm}

\sloppy
\thispagestyle{empty}
\begin{flushleft}
DESY 18--196, 
\\
DO--TH 19/31\\
TTP 19--045\\
SAGEX 19--34\\
\end{flushleft}

\mbox{}
\vspace*{\fill}
\begin{center}

{\LARGE\bf The \boldmath $O(\alpha^2)$ Initial State QED Corrections} 

\vspace*{2mm}
{\LARGE\bf to \boldmath $e^+e^- \rightarrow \gamma^*/Z_0^*$}

\vspace{3cm}
\large
{\large 
J.~Bl\"umlein$^a$, 
A.~De~Freitas$^a$,
C.G.~Raab$^b$,
and
K.~Sch\"onwald$^{a,c}$
}

\vspace{1.cm}
\normalsize
{\it   $^a$~Deutsches Elektronen--Synchrotron, DESY,}\\
{\it   Platanenallee 6, D--15738 Zeuthen, Germany}

\vspace*{2mm}
{\it  $^b$~Institut f\"ur Algebra, Johannes Kepler Universit\"at Linz,}\\
{\it Altenberger Stra\ss{}e 69, A--4040 Linz, Austria}

\vspace*{2mm}
{\it  $^c$~Institut f\"ur Theoretische Teilchenphysik, \\
Karlsruher Institut f\"ur Technologie (KIT) D-76128 Karlsruhe, Germany}


\end{center}
\normalsize
\vspace{\fill}
\begin{abstract}
\noindent
We calculate the complete $O(\alpha^2)$ initial state radiation corrections to $e^+ e^-$ annihilation into 
a neutral vector boson in a direct analytic computation without any approximation. The corrections are 
represented in terms of iterated incomplete (elliptic) integrals over alphabets of square--root valued 
letters. Performing the limit $s \gg m_e^2$, we find discrepancies with the earlier results of 
Ref.~\cite{Berends:1987ab} and confirm results obtained in Ref.~\cite{Blumlein:2011mi} where the effective 
method of massive operator matrix elements has been used, which works for all but the power corrections in 
$m^2_e/s$. In this way, we also confirm the validity of the factorization of massive partons in the Drell--Yan 
process. We add non--logarithmic terms at $O(\alpha^2)$ which have not been considered in previous calculations. 
The final results in the limit $s \gg m_e^2$ can be given in terms of Nielsen integrals. 
\end{abstract}

\vspace*{\fill}
\noindent
\newpage 

\section{Introduction}
\label{sec:1}

\vspace*{1mm}
\noindent
$e^+e^-$ colliders operating at high energy and at large luminosity  measure the fundamental parameters of the Standard
Model with high precision and  perform crucial tests on the structure of the Standard Model. In the past the experiments at
LEP obtained very precise results on the parameters of the $Z$-boson \cite{ALEPH:2005ab}. The future large scale 
facilities like the ILC, CLIC \cite{ILC,Aihara:2019gcq,Mnich:2019,CLIC}, the FCC\_ee \cite{FCCEE}, and muon colliders 
\cite{Delahaye:2019omf} are planned to operate at even higher energies and luminosities. There one can perform in 
addition also precise scans of the $t\bar{t}$-threshold measuring the properties of the top quark in detail and produce 
the Higgs boson under very clean conditions in $ZH$-final states, which will finally allow to understand more properties 
of the Higgs boson in great detail.

One important condition to perform these highly precise measurements is the exact knowledge of the QED radiative 
corrections for the process $e^+e^- \rightarrow \gamma^*/Z^*$, which has to be known to two--loop order in the 
fine structure constant $\alpha$, adding further logarithmic contributions in higher orders. A first calculation 
of the $O(\alpha^2)$ initial state radiative corrections to this process has been performed in 
Ref.~\cite{Berends:1987ab}. In this reference various approximations have been made in the integrands of the 
Feynman 
diagrams, to simplify the integration process. In 2011 it has been noticed, however, in a second calculation 
based on massive operator matrix elements (OMEs) \cite{Blumlein:2011mi} that the results deviated in all 
channels for the constant term at $O(\alpha^2)$, while the $O(\alpha)$ result and the logarithmic terms at 
$O(\alpha^2)$ agreed. In the latter calculation it was assumed that the Drell--Yan process with massive 
external lines factorizes. At that time, the new results 
did not yield a thorough counter argument against the results in \cite{Berends:1987ab}, since one might have 
argued that there is no factorization in the massive Drell--Yan process.

There is actually only one way to decide which of the results is correct. One has to perform the complete calculation of 
the scattering cross section without any approximation or assumption analytically. In the final result one will of course 
expand in the ratio $m_e^2/s \approx 3 \cdot 10^{-11}$, where $m_e$ denotes the electron mass and $s$ the cms energy 
squared to obtain a compact result.

In Ref.~\cite{Berends:1987ab} some processes which only contribute to the $O(\alpha^2)$ term and have no logarithmic 
contributions were not considered. A first calculation, however, in the massless case, was performed in \cite{Hamberg:1990np} 
and later in \cite{Harlander:2002wh} for the Drell--Yan process in Quantum Chromodynamics (QCD). Furthermore, 
differences appearing in 
the calculation of the contributing vector and axial--vector terms were not considered in \cite{Berends:1987ab}.

In the present paper we perform a thorough analytic calculation of all contributing terms. 
We confirm the results given in \cite{Blumlein:2011mi} before and we add the pure $O(\alpha^2)$ terms, which cannot 
be derived
using the method of massive OMEs \cite{Buza:1995ie}. The calculation in Ref.~\cite{Blumlein:2011mi} has been performed for 
vector couplings. Here we add also the axial--vector contributions, whenever they are not suppressed by power 
corrections of $O(m_e^2/s)$. Our final results are expanded in $m_e^2/s$ and we maintain all terms up to 
$O((m_e^2/s)^0)$.
The final radiators can be expressed by Nielsen integrals \cite{NIELSEN}
\begin{eqnarray}
S_{p,n}(x)   &=& \frac{(-1)^{n+p-1}}{(n-1)! p!} \int_0^1 \frac{dt}{t} \ln^{n-1}(t) \ln^p(1-zt),\\
S_{n-1,1}(x) &=& \Li_n(x), 
\end{eqnarray}
which cover the classical polylogarithms \cite{CLPOLY}.

The radiator functions have the general structure
\begin{eqnarray}
R\left(z,\alpha,\frac{s}{m^2}\right) &=& \delta(1-z) + \sum_{k=1}^\infty \left(\frac{\alpha}{4\pi}\right)^k
C_k\left(z, \frac{s}{m^2}\right) \\ C_k\left(z, \frac{s}{m^2}\right) &=& \sum_{l=0}^k \ln^{k-l}\left(\frac{s}{m^2}\right)
c_{k,l}(z).
\end{eqnarray}
The respective differential cross sections are then given by
\begin{eqnarray}
\frac{d \sigma_{e^+e^-}}{ds'} = \frac{1}{s} \sigma_{e^+e^-}(s') R\left(z,\alpha,\frac{s}{m^2}\right),
\end{eqnarray}
with $\sigma_{e^+e^-}(s')$ the scattering cross section without the initial state radiation (ISR) corrections and  
\begin{eqnarray}
z = \frac{s'}{s}, 
\end{eqnarray}
where $s'$ is the invariant mass of the produced (off-shell) $\gamma/Z$ boson. Here and in the following the mass $m$ denotes
the electron mass $m_e$, if not stated otherwise.

The paper is organized as follows. In Section~\ref{sec:2} we present the Born cross section for the process.
The $O(\alpha)$ corrections are given in Section~\ref{sec:3}. General aspects of the integration at two--loop order
are discussed in Section~\ref{sec:4}. In Section~\ref{sec:5} we present the results for the different processes 
contributing to the two--photon corrections. The non--singlet process of the $e^+e^-$ pair radiation 
process is discussed in Section~\ref{sec:6}. The contribution due to the 
radiation of heavier final states in the non--singlet process is calculated in Section~\ref{sec:7}, followed 
by those due to the pure--singlet process,
Section~\ref{sec:8}, and the interference term between the non--singlet and the pure singlet terms, Section~\ref{sec:9}, 
in the case of $e^+e^-$ emission. In Section~\ref{sec:10}, we give the results for processes that have no 
logarithmic contributions at $O(\alpha^2)$. 
The axial--vector contributions are discussed for the
processes they contribute to. In all other cases the radiators are the same as in the vector case. 
Finally,  we discuss the soft--photon exponentiation  contributions beyond the radiative 
corrections to $O(\alpha^2)$ in Section~\ref{sec:11}, and Section~\ref{sec:12} contains the conclusions. Numerical 
results 
on the $Z$ peak, for the $ZH$ production process and $t\overline{t}$ production have already been presented in 
Ref.~\cite{Blumlein:2019pqb}. There and in \cite{Blumlein:2019srk} we also line out the numerical differences to 
\cite{Berends:1987ab}. In Appendix~\ref{sec:A}  we present details on phase--space integrals which have been performed 
in the present paper. 
\section{The Process}
\label{sec:2}

\vspace*{1mm}
\noindent
We consider the process of $e^+e^-$ annihilation into a virtual photon $\gamma^*$ or virtual $Z_0^*$ boson above 
a mass
threshold of $s' \geq 4 m_\mu^2$ or larger, with $m_\mu$ the muon mass and $s$ the cms energy squared of the 
annihilation process. Also the production of other fermionic final states can be considered such as $\tau^+\tau^-$,
massless $q\bar{q}$ and the corresponding heavy quark pairs. The phase space limit on $s'$ is $s' \geq 4 m_f^2$.
We will usually assume $s' \geq 4 m_\mu^2$ or a more conservative cut.
\begin{figure}[th]
  \centering
  \hskip-0.8cm
  \includegraphics[width=.3\linewidth]{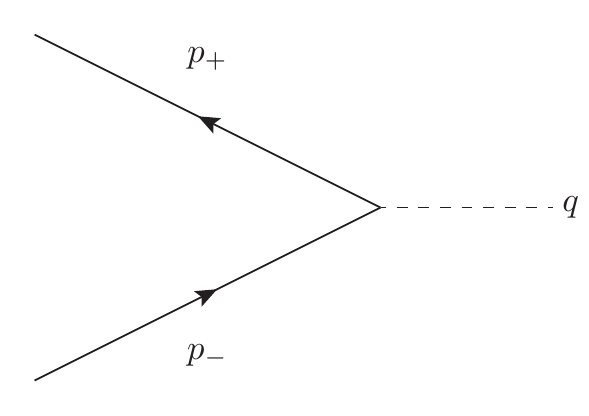} \hspace*{1cm}
\caption{\sf The Born cross section for the process $e^+e^- \rightarrow Z^*/\gamma^*$.}
  \label{FIG1}
\end{figure}
The differential Born cross section is given by
\begin{eqnarray}
\frac{d \sigma_{e^+ e^-}^{(0),\rm I}}{d s'} = \delta(s-s') \sigma^{(0)}(s'),
\end{eqnarray}
where $\sigma^{(0)}(s')$ denotes the integrated cross section of one of the above processes. It corresponds to the 
annihilation diagram in Figure~\ref{FIG1}.
For $s$-channel $e^+e^-$ annihilation into a virtual gauge boson $(\gamma^*, Z^*)$, which decays into a fermion 
pair $f \bar{f}$, the scattering cross section reads
\begin{eqnarray}
\label{eq:BO1}
\frac{d\sigma^{(0)}(s)}{d \Omega} &=& \frac{\alpha^2}{4 s}
                                        N_{C,f} \sqrt{1 - \frac{4 m_f}{s}}
\nonumber\\
& & \times
\left[\left(1+ \cos^2\theta + \frac{4 m_f^2}{s} \sin^2\theta \right) G_1(s)
- \frac{8 m_f^2}{s} G_2(s)
+ 2 \sqrt{1-\frac{4m_f^2}{s}} \cos\theta G_3(s)\right]
\nonumber\\ && \times {\cal G}(s)~,
\\
\label{eq:BO2}
\sigma^{(0)}(s) &=& \frac{4 \pi \alpha^2}{3 s}
                                        N_{C,f} \sqrt{1 - \frac{4 m_f}{s}}
\left[\left(1 + \frac{2 m_f^2}{s} \right) G_1(s)
- 6 \frac{m_f^2}{s} G_2(s)\right] {\cal G}(s)~,
\end{eqnarray}
see e.g. \cite{BDJ,WB}.\footnote{Note a missing term in \cite{Berends:1987ab}, Eq.~(2.5).}
Here the
final state fermions are considered not to be electrons, to obtain an $s$-channel Born cross section. In 
Eqs.~(\ref{eq:BO1}, \ref{eq:BO2}) the electron mass is neglected kinematically. $\alpha$ denotes the fine structure 
constant, $N_{C,f}$ is the number of colors of the final state fermion, with $N_{C,f} = 1$ for colorless fermions, 
and $N_{C,f} = 3$ for quarks. The function ${\cal G}(s)$ is set to $1$ in the case of the pure perturbative 
calculation.
$s$ is the cms energy, $\Omega$  the spherical angle, $\theta$ the cms scattering 
angle, and the effective couplings $G_i(s)|_{i=1...3}$ read
\begin{eqnarray}
G_1(s) &=& Q_e^2 Q_f^2 + 2 Q_e Q_f v_e v_f {\sf Re}[\chi_Z(s)]
          +(v_e^2+a_e^2)(v_f^2+a_f^2)|\chi_Z(s)|^2,\\
G_2(s) &=& (v_e^2+a_e^2) a_f^2 |\chi_Z(s)|^2, \\
G_3(s) &=& 2 Q_e Q_f a_e a_f {\sf Re}[\chi_Z(s)] + 4 v_e v_f a_e a_f |\chi_Z(s)|^2.
\end{eqnarray}
The reduced $Z$--propagator is given by
\begin{eqnarray}
\chi_Z(s) = \frac{s}{s-M_Z^2 + i M_Z \Gamma_Z},
\end{eqnarray}
where $M_Z$ and  $\Gamma_Z$ are the mass and the with of the $Z$--boson and $m_f$ is the mass of the final
state fermion. $Q_{e,f}$ are the electromagnetic charges of the electron $(Q_e = -1)$
and the final state fermion, respectively, and
the electro--weak couplings $v_i$ and $a_i$ are given by
\begin{eqnarray}
v_e &=& \frac{1}{\sin\theta_w \cos\theta_w}\left[I^3_{w,e} - 2 Q_e
\sin^2\theta_w\right],\\
a_e &=& \frac{1}{\sin\theta_w  \cos\theta_w} I^3_{w,e}, \\
v_f &=& \frac{1}{\sin\theta_w \cos\theta_w}\left[I^3_{w,f} - 2 Q_f
\sin^2\theta_w\right],\\
a_f &=& \frac{1}{\sin\theta_w  \cos\theta_w} I^3_{w,f}~,
\end{eqnarray}
where $\theta_w$ is the weak mixing angle, and $I^3_{w,i} = \pm 1/2$ the third component
of the weak isospin for up and down particles, respectively.

For the radiative corrections studied below, we will consider the integrated cross section (\ref{eq:BO2}) in the energy region 
of the $Z$--peak. 

In the following we will use the fine structure constant with the normalization
\begin{eqnarray}
a = \frac{\alpha}{4 \pi}.
\end{eqnarray}
The scattering cross section including the contributions due to initial state radiation can be expressed as follows
\begin{eqnarray}
\sigma(s) = \int_0^1 dz R(z;a,L) \sigma_0(z s),
\end{eqnarray}
where $R(z;a,L)$ is the distribution--valued \cite{DISTR} radiation function, with 
\begin{eqnarray}
L = \ln\left(\frac{s}{m_e^2}\right).
\end{eqnarray}
The different radiators calculated in the present paper sum to the following distribution
\begin{eqnarray}
R(z;a,L) 
&=&  a R_1^\gamma(z,L) + a^2 \bigl[
R_2^{\gamma\gamma}(z,L) + 
R_2^{e^+e^-,\rm NS}(z,L) +  
R_2^{f \bar{f},\rm NS}(z,L) +  
R_2^{e^+e^-,\rm PS}(z,L) 
\nonumber \\ &&
+  
R_2^{{e^+e^-,\rm NS}-{\rm PS~interf.}}(z,L) +  
R_2^{e^+e^-,\rm non log}(z,L) \bigr]
+ R_{\rm soft, 3}(z;a,L). 
\end{eqnarray}
\section{The One--Loop Corrections}
\label{sec:3}

\vspace*{1mm}
\noindent
At one--loop order only photonic corrections contribute. We will work in $D = 4$ dimensions. This allows to treat axial--vector
couplings without an additional finite renormalization. 
\begin{figure}[H]
  \centering
  \hskip-0.8cm
\hspace*{1cm}
  \includegraphics[width=.8\linewidth]{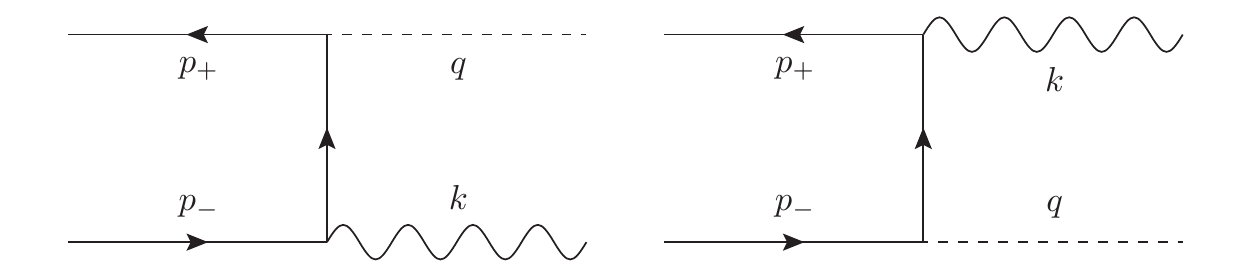}
  \caption{\sf The $O(\alpha)$ $e^+e^-$ annihilation graphs into a photon and a virtual gauge boson. 
}
  \label{fig:h1}
\end{figure}
\begin{figure}[H]
  \centering
  \hskip-0.8cm
  \includegraphics[width=.7\linewidth]{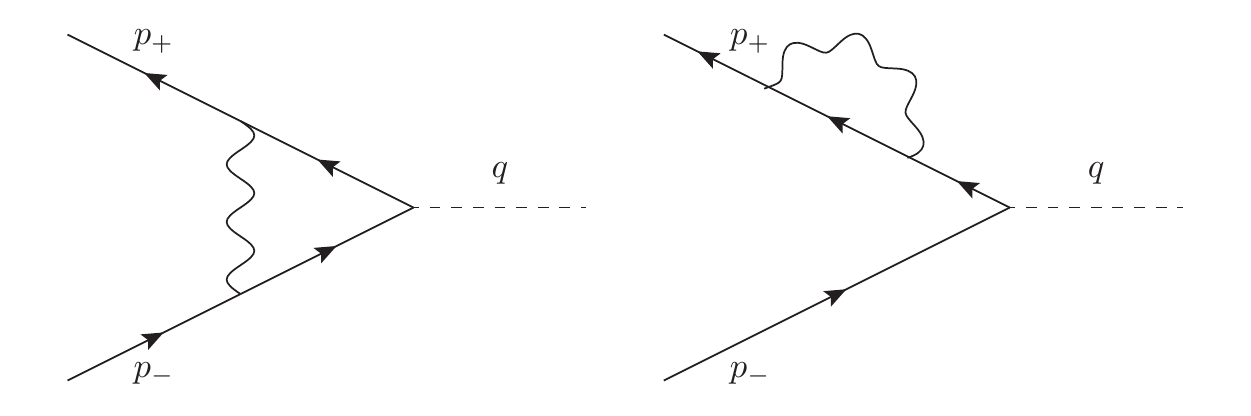} 
  \caption{\sf The $O(\alpha)$ virtual corrections  to $e^+e^-$ annihilation into virtual gauge boson.}
  \label{fig:v1}
\end{figure}
The Dirac algebra was performed
using {\tt FORM} \cite{Vermaseren:2000nd}. In the integration of the scalar integrals also the packages  
{\tt Sigma} \cite{Schneider:2007a,Schneider:2013a}, {\tt HarmonicSums},
\cite{Vermaseren:1998uu,Blumlein:1998if,Ablinger:2014rba,Ablinger:2010kw, Ablinger:2013hcp, Ablinger:2011te, 
Ablinger:2013cf,Ablinger:2014bra,Ablinger:2017Mellin} were used. Later, for finding representations of iterated 
integrals over
special alphabets we used also  the package {\tt HolonomicFunctions} \cite{KOUTSCHAN}, and private implementations 
\cite{RAAB1}. 

The photon radiation and virtual diagrams are given in Figures~\ref{fig:h1} 
and \ref{fig:v1}, respectively.
We parameterize the different contributions by
\begin{eqnarray}
\label{eq:SIG1}
R_1^\gamma =
\delta(1-z) 
\left[
T_1^{\rm S_1}(\lambda,m_e,L,\ep) + T_1^{\rm V_1}(\lambda,m_e,L)\right] + \theta(1-z-\ep) T_1^{\rm H_1}(L,z),
\nonumber\\
\end{eqnarray}
where the indices S, V, and H stand for soft, virtual and hard. Let $\Delta$ denote the energy cut--off for 
soft--photon bremsstrahlung. Then 
\begin{eqnarray}
\ep = \frac{2 \Delta}{s} \ll 1.
\end{eqnarray}
Here
$\lambda$ denotes a soft--photon mass, with $\lambda \ll m_e$. This parameter can be introduced in an Abelian gauge
theory calculation and it cancels in the final result. 

One obtains 
\begin{eqnarray}
T_1^{\rm S_1}(\lambda,m_e,L,\ep) = 4 \left [ \frac{1}{2} L^2 - \ln\left(\frac{\lambda^2}{m^2_e}\right) (L-1) + 2 \ln(\ep) 
(L-1)- 2 \zeta_2\right],
\end{eqnarray}
and $\zeta_k = \sum_{l=1}^\infty (1/l^k),~~k \in \mathbb{N}, k \geq 2$ is the Riemann $\zeta$-function at integer
argument.

The virtual corrections are obtained from the one-loop massive Dirac form factor in the limit $s \gg m_e^2$
\begin{eqnarray}
T_1^{\rm V_1}(\lambda,m_e,L) = 2 {\sf Re} F^{(1)}(s), 
\end{eqnarray}
with \cite{Bonneau:1971mk,Barbieri:1972hn,Barbieri:1972as}
\begin{eqnarray}
\label{eq:F1}
F^{(1)}(s) &=&   - L^2 +2 \ln\left(\frac{\lambda^2}{m^2_e}\right)(L-1) + 3 L - 4 + 8 \zeta_2
+ i\pi \left( 2 L  - 2  \ln\left(\frac{\lambda^2}{m^2_e}\right) -3 \right) .
\end{eqnarray}
The Pauli form factor is power suppressed in the limit $s \gg m_e^2$.
Finally, the hard contribution is given by
\begin{eqnarray}
T_1^{\rm H_1}(L,z) = 4 \frac{1+z^2}{1-z} (L-1).
\end{eqnarray}
Here there is also a lower bound on $z$ from the minimal value of $s' \geq 4 m_f^2$. One obtains for (\ref{eq:SIG1})
the well--known result
\begin{eqnarray}
R_1^\gamma = 
P_{ee}^{(0)}(z) (L-1) \theta(1-z-\varepsilon) + 2\bigl(4\zeta_2 -4 + 3 L + 4(L-1)\ln(\varepsilon) \bigr) \delta(1-z),
\end{eqnarray}
cf.~\cite{Berends:1987ab}, \cite{Blumlein:2011mi}, Eqs.~(40, 96), where 
\begin{eqnarray}
P_{ee}^{(0)}(z) &=& 4 \frac{1+z^2}{1-z}
\end{eqnarray}
denotes the first order electron--electron splitting function.
It can be promoted to a $+$-distribution where the $+$-prescription is defined by
\begin{eqnarray}
\label{eq:PLUS}
\int_0^1 dx \left[f(x)\right]_+ g(x) = \int_0^1 dx f(x) \left[g(x)-g(1)\right].
\end{eqnarray}
Then one can drop the $\theta$-function and all contributions proportional to $\ln(\varepsilon)$.

The above relations were derived for the vector case. The same corrections are obtained in the axial--vector
case in the limit $s \gg m_e^2$ since the difference at one--loop order is suppressed by power corrections 
in $m_e^2/s$.
\section{Analytic Integration of the \boldmath $O(\alpha^2)$ Corrections}
\label{sec:4}

\vspace*{1mm}
\noindent
For the $O(\alpha^2)$ terms, non--trivial phase space integrals are occurring given by fourfold integrals. 
Details
on their calculation are presented in Appendix~\ref{sec:A}. They are given by two angular integrals and two further 
integrals over invariants. In course of these integrations one obtains square--root valued arguments in logarithms and  
polylogarithms, which are nested in part and have to be rationalized or transformed to single roots to perform the next 
integration.\footnote{For an algorithmic approach to rationalization see \cite{Besier:2018jen,Besier:2019kco}.} 
The principal way to obtain the 
corresponding square root--valued iterated integrals, containing real 
parameters in the letters, has been described already in Refs.~\cite{Blumlein:2019qze,Blumlein:2019zux}. Square
root--valued iterated integrals based on rational parameters have been considered earlier in \cite{Ablinger:2014bra}.
They occur as Mellin inversions of finite binomial and inverse binomial sums.

In total, up to weight {\sf w = 3} iterated integrals emerge. We aim on an analytic iterated integral representation 
over an alphabet also containing square root--valued letters and will keep the complete dependence on 
\begin{eqnarray}
\rho = \frac{m_e^2}{s}.
\end{eqnarray}
Using special variables, it is also possible to expand in $\rho \ll 1$ prior to the last integration is carried out.
The integration has been performed using {\tt Mathematica}. After this expansion one obtains a large number of 
logarithms and classical polylogarithms $\Li_2(g_i(z)), \Li_3(g_i(z))$ with involved, partly complex arguments. 
They have to be mapped to logarithms and polylogarithms of the convenient arguments $z$ and $1-z$. For this we 
use associated differential equations. Some of the polylogarithms also depend on $G$-functions 
\cite{Ablinger:2014bra,Ablinger:2017err,Ablinger:2017xml,Ablinger:2018brx}, containing square root--valued letters. 
Here the $G$-functions are defined by
\begin{eqnarray}
G(\{b,\vec{a}\},x) = \int_0^x h(\{b\},x)
  G(\{\vec{a}\},x),~~~G(\{\emptyset\},x)
  = 1,~~h(\{c_i\},x) \in \mathfrak{A}',
\end{eqnarray}
over an alphabet $\mathfrak{A}'$. The different letters $h(\{c_i\},x)$ are not yet independent w.r.t. the 
associated differential field.
Adding all contributions, the $G$-functions cancel.

We finally compare the exact analytic result, not expanded in $\rho$, with the expanded result including the 
$O(\rho^0)$ terms, numerically. The expansion in $\rho$ can also be performed starting with the complete result. 
This requires the introduction of suitable regularizations. We have performed the last step in the non--singlet case 
and obtained the same
result as expanding below the last integral, using appropriate variables. In all cases the numerical comparison shows 
a relative agreement of $O(10^{-7})$ at $s= M_Z^2$, which is the expected result in this approximation. It shows that
the formulae expanded in the light fermion mass can be used for experimental analyses.

The complete analytic results will be given in terms of iterative integrals over a certain alphabet of
letters $\mathfrak{A}$, which are mostly square--root valued and contain real parameters. Some of them were 
occurring already in earlier
investigations \cite{Ablinger:2014bra}. They are labeled by $v_i$. Other letters are new and are named by $d_i$.
The iterative integrals are given by
\begin{eqnarray}
H^*_{b,\vec{a}}(x) = \int_x^1 dy f_{b}(y) H^*_{\vec{a}}(y),~~~~H^*_\emptyset = 1.
\end{eqnarray}
The different letters of the alphabet, $f_{k}(t;z,\rho) \equiv f_k$ are~:
\begin{align}
	d_1 &= \frac{1}{ \sqrt{1-t} \sqrt{16 \rho^2 - 8 \rho (1+z) t + (1-z)^2 t^2} } ,
\\
	d_2 &= \frac{t}{ \sqrt{1-t} \sqrt{16 \rho^2 - 8 \rho (1+z) t + (1-z)^2 t^2} } ,
\\
	d_3 &= \frac{1}{t \sqrt{1-t} \sqrt{16 \rho^2 - 8 \rho (1+z) t + (1-z)^2 t^2} } ,
\\
	d_4 &= \frac{1}{ \bigl( 16 \rho^2 + ( 4 z - 8 \rho ( 1 + z ) ) t + (1-z)^2 t^2 \bigr)  \sqrt{1-t} \sqrt{16 \rho^2 - 8 \rho (1+z) t + (1-z)^2 t^2} } ,
\\
	d_5 &= \frac{t}{ \bigl( 16 \rho^2 + ( 4 z - 8 \rho ( 1 + z ) ) t + (1-z)^2 t^2 \bigr) \sqrt{1-t} \sqrt{16 \rho^2 - 8 \rho (1+z) t + (1-z)^2 t^2} } ,
\\
	d_6 &= \frac{1}{ \bigl( 16 \rho^2 + ( 4 z - 8 \rho ( 1 + z ) ) t + (1-z)^2 t^2 \bigr) \sqrt{16 \rho^2 - 8 \rho (1+z) t + (1-z)^2 t^2} } ,
\\
	d_7 &= \frac{t}{ \bigl( 16 \rho^2 + ( 4 z - 8 \rho ( 1 + z ) ) t + (1-z)^2 t^2 \bigr) \sqrt{16 \rho^2 - 8 \rho (1+z) t + (1-z)^2 t^2} } ,
\\
	d_8 &= \frac{1-z}{\bigl( 4 \rho - (1-z) t \bigr) \sqrt{1-t} } ,
\\
	d_9 &= \frac{1}{ \bigl( 16 \rho^2 + 4 \bigl( z - 2 \rho ( 1 + z ) \bigr) t + (1-z)^2 t^2 \bigr) \sqrt{1-t} },
\\
	d_{10} &= \frac{t}{ \bigl( 16 \rho^2 + 4 \bigl( z - 2 \rho ( 1 + z ) \bigr) t + (1-z)^2 t^2 \bigr) \sqrt{1-t} },
\\
	d_{11} &= \frac{1}{t \sqrt{16 \rho^2 - 8 \rho (1+z) t + (1-z)^2 t^2} },
\\
	d_{12} &= \frac{1}{16 \rho^2 + 4 \bigl( z - 2 \rho ( 1 + z ) \bigr) t + (1-z)^2 t^2},
\\
	d_{13} &= \frac{t}{16 \rho^2 + 4 \bigl( z - 2 \rho ( 1 + z ) \bigr) t + (1-z)^2 t^2},
\\
      d_{14} &= \frac{1}{t(1-z) -4 \rho },
\\
      d_{15} &= \frac{1}{\sqrt{1-t} (t(1-z)-4 \rho )},
\\
      d_{16} &= \frac{1}{\sqrt{t (1-t)} \sqrt{t (1-z)^2-16 \rho ^2}},
\\
      d_{17} &= \frac{1}{\sqrt{t(1-t)} (t (1-z)-4 \rho ) \sqrt{t (1-z)^2-16 \rho ^2}},
\\
      d_{18} &= \frac{1}{\sqrt{t} \sqrt{t (1-z)^2-16\rho ^2}},
\\
      d_{19} &= \frac{1}{\sqrt{t} (t (1-z)-4 \rho ) \sqrt{t (1-z)^2-16 \rho ^2}},
\\
      d_{20} &= \frac{1}{\sqrt{t^2 (1-z)^2-8 \rho  t (1+z)+16 \rho ^2}},
\\
      d_{21} &= \frac{1}{\sqrt{1-t} \sqrt{t^2 (1-z)^2-8 \rho  t (1+z)+16 \rho ^2}},
\\
      d_{22} &= \frac{\sqrt{t}}{\sqrt{t (1-z)^2-16 \rho ^2} \sqrt{t^2 (1-z)^2-8 \rho  t (1+z)+16 \rho ^2}},
\\
      d_{23} &= \frac{\sqrt{t}}{\sqrt{t (1-z)^2-16 \rho ^2} \big(t^2(1-z)^2 - 8 \rho(1+z)t + 4 t z + 16 \rho^2 \big)},
\\
      d_{24} &= \frac{1}{\big(t^2(1-z)^2 - 8 \rho(1+z)t + 4 t z + 16 \rho^2 \big) \sqrt{t^2 (1-z)^2-8 \rho  t (1+z)+16 \rho ^2}},
\\
      d_{25} &= \frac{t}{\big(t^2(1-z)^2 - 8 \rho(1+z)t + 4 t z + 16 \rho^2 \big) \sqrt{t^2 (1-z)^2-8 \rho  t (1+z)+16 \rho ^2}},
\\
      d_{26} &= \frac{1}{\sqrt{1-t} \big(t^2(1-z)^2 - 8 \rho(1+z)t + 4 t z + 16 \rho^2 \big) \sqrt{t^2 (1-z)^2-8 \rho  t (1+z)+16 \rho ^2}},
\\
      d_{27} &= \frac{t}{\sqrt{1-t} \big(t^2(1-z)^2 - 8 \rho(1+z)t + 4 t z + 16 \rho^2 \big) \sqrt{t^2 (1-z)^2-8 \rho  t (1+z)+16 \rho ^2}},
\\
      d_{28} &= \frac{1}{\sqrt{t}\sqrt{t (-1+z)^2-16 \rho ^2} \sqrt{t^2 (1-z)^2-8 \rho  t (1+z)+16 \rho ^2}},
\\
      d_{29} &= \frac{1}{\sqrt{t} \sqrt{t (1-z)^2-16 \rho ^2} \big(t^2(1-z)^2 - 8 \rho(1+z)t + 4 t z + 16 \rho^2 \big)},
\\
      d_{30} &= \frac{1}{\sqrt{t} \sqrt{t (1-z)^2-16 \rho ^2} \big(t^2(1-z)^2 - 8 \rho(1+z)t + 4 t z + 16 \rho^2 )} \nonumber\\
& \times  \frac{1}{\sqrt{t^2(1-z)^2 - 8 \rho(1+z)t + 16 \rho^2}},
\\
      d_{31} &= \frac{\sqrt{t}}{\sqrt{t (1-z)^2-16 \rho ^2} \big(t^2(1-z)^2 - 8 \rho(1+z)t + 4 t z + 16 \rho^2 \big) \sqrt{t^2 (1-z)^2-8 \rho  t (1+z)+16 \rho ^2}},
\\
      d_{32} &= \frac{1}{t \sqrt{1-t} \sqrt{t^2 (1-z)^2-8 \rho  t (1+z)+16 \rho ^2}} ,
\\
      d_{33} &= \frac{t}{\sqrt{1-t} \sqrt{t^2 (1-z)^2-8 \rho  t (1+z)+16 \rho ^2}},
\\
	v_1 &= \frac{1}{ \sqrt{1-4t} \sqrt{ 16 t^2 - 8 (1+z) t + (1-z)^2} },
\\ 
	v_2 &= \frac{1}{ t \sqrt{1-4t} \sqrt{ 16 t^2 - 8 (1+z) t + (1-z)^2} },
\\
	v_3 &= \frac{1}{ \sqrt{1-4t} \bigl( 4t - (1+x) \bigr) \sqrt{ 16 t^2 - 8 (1+z) t + (1-z)^2} },
\\
	v_4 &= \frac{1}{t \sqrt{1-t}} .
\end{align}
We call the iterated integrals containing both Kummer--type letters \cite{Ablinger:2013cf} and the above letters,
Kummer--elliptic integrals, since some of them integrate to elliptic structures, although they do so as indefinite 
integrals,\footnote{For related alphabets occurring in other calculations see e.g.~\cite{Adams:2018kez,Becchetti:2019tjy}.}
which are still iterative if compared to complete elliptic integrals and their extensions, cf.~\cite{ELLIPTIC}.

We label the processes using the same scheme as in Ref.~\cite{Berends:1987ab}, i.e. process I: the photon emission case;
process II: the non--singlet case for fermion-pair production; process III: the pure singlet process; process IV:
the interference term between the non--singlet case for $e^+e^-$ pair emission and the pure singlet case. Furthermore,
we denote contributions not covered in \cite{Berends:1987ab} but belonging to the $O(a^2)$ QED corrections as process B,
in accordance with Ref.~\cite{Hamberg:1990np}.

A few remarks on the size of the present calculation are in order.  {\it i)} the size of the amplitudes amounts to 
10~Gb (process I),
25~kb (process II),
56~kb (process III) and
124~kb (process IV). The calculation of process I required several months of code design and 30h of 
computation 
time.
The reduction to the basis of iterative integrals took 1 day (process II), 1 month (process III), and 2 months (process 
IV). The integration time for processes II--IV amounted to minutes, 2h and 5h. No essential resources were 
necessary to perform the calculation for process B.
The size of the project mainly resulted
from the fact that only in the last step an expansion in the parameter $\rho$ has been performed. The required
computer power was not available at the time when Ref.~\cite{Berends:1987ab} was worked out. This also 
applies to several computer-algebraic and mathematical calculation techniques we were able to apply, which 
became available only recently.

Now we turn to the calculation of the individual sub--processes at two--loop order.
\section{The Photonic Two-Loop Corrections}
\label{sec:5}

\begin{figure}[H]
  \centering
  \hskip-0.8cm
  \includegraphics[width=.25\linewidth]{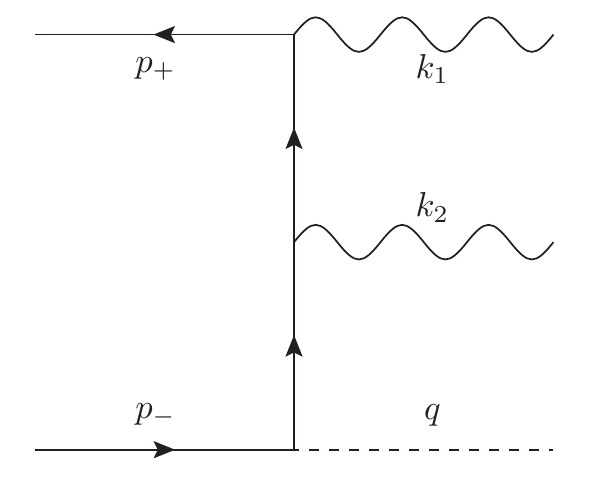}
  \includegraphics[width=.25\linewidth]{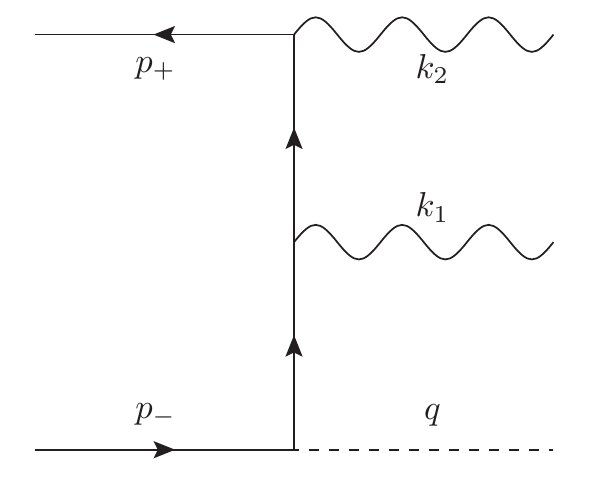}
  \includegraphics[width=.25\linewidth]{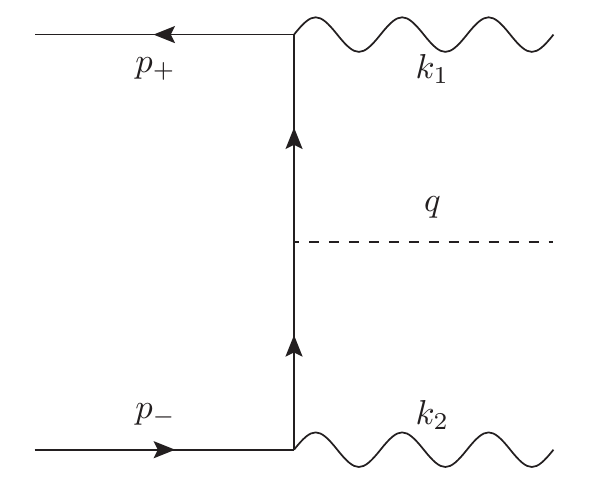}
  \includegraphics[width=.25\linewidth]{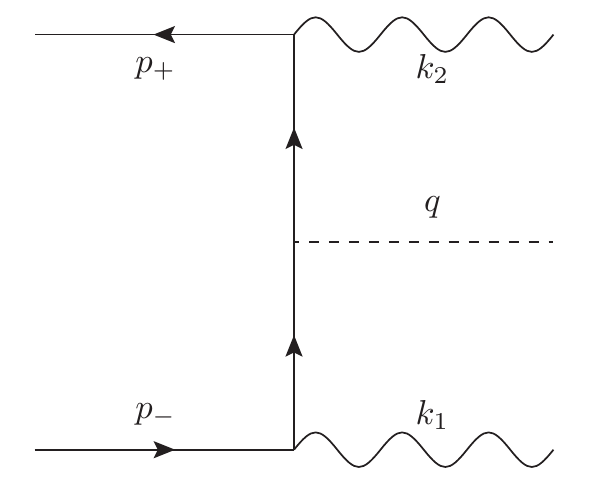}
  \includegraphics[width=.25\linewidth]{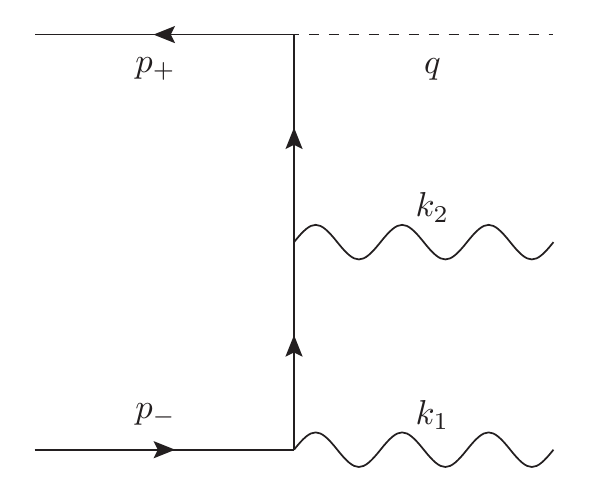}
  \includegraphics[width=.25\linewidth]{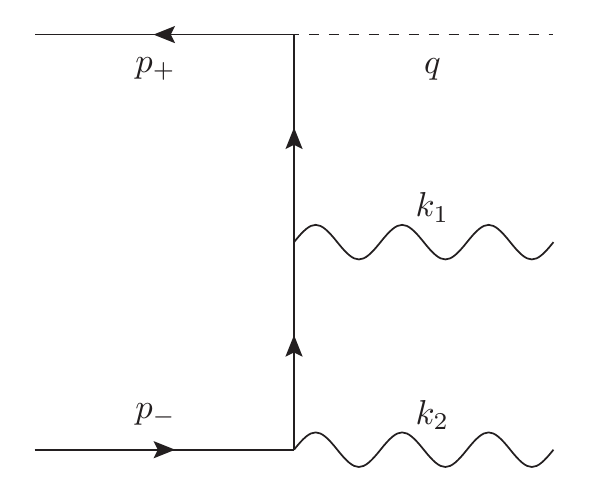}
  \caption{\sf The $e^+e^-$ annihilation graphs into  two photons and a virtual gauge boson.}
  \label{fig:gaga}
\end{figure}

\begin{figure}[H]
  \centering
  \hskip 0.1cm
  \includegraphics[width=.24\linewidth,trim={0 11.5cm 2.6cm 0}]{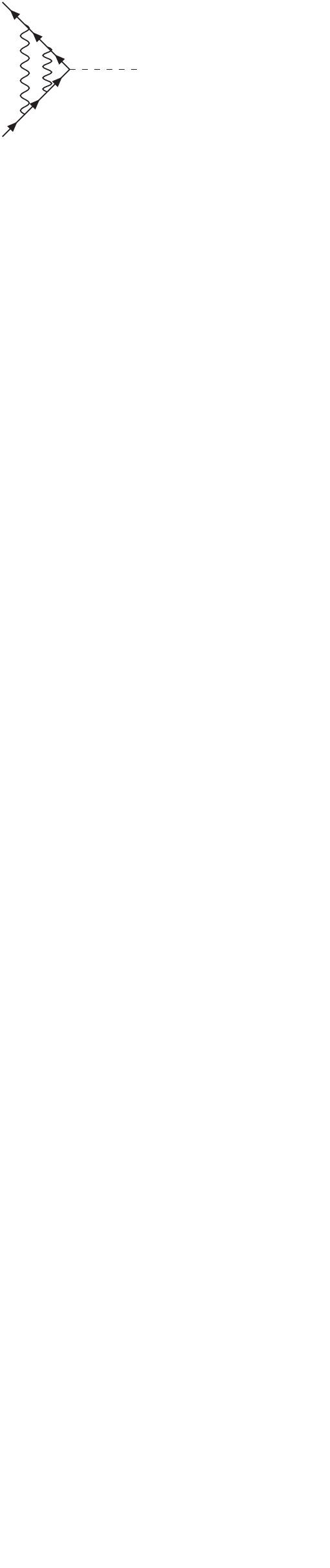}
  \includegraphics[width=.24\linewidth,trim={0 11.5cm 2.6cm 0}]{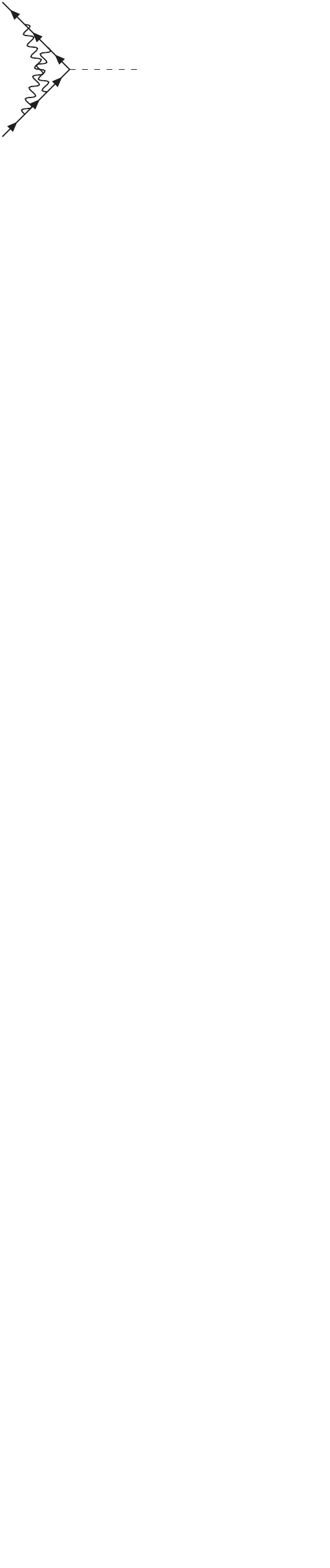}
  \includegraphics[width=.24\linewidth,trim={0 11.5cm 2.6cm 0}]{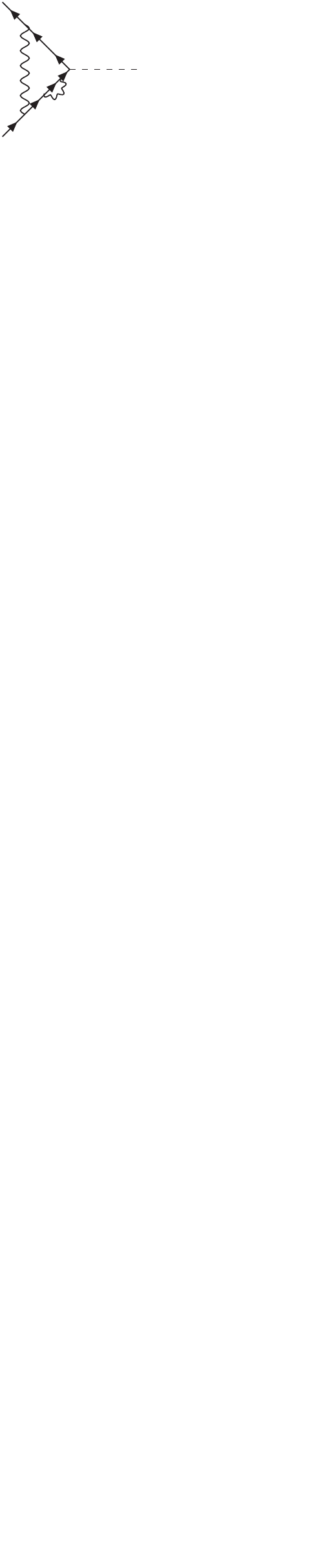}
  \includegraphics[width=.24\linewidth,trim={0 11.5cm 2.6cm 0}]{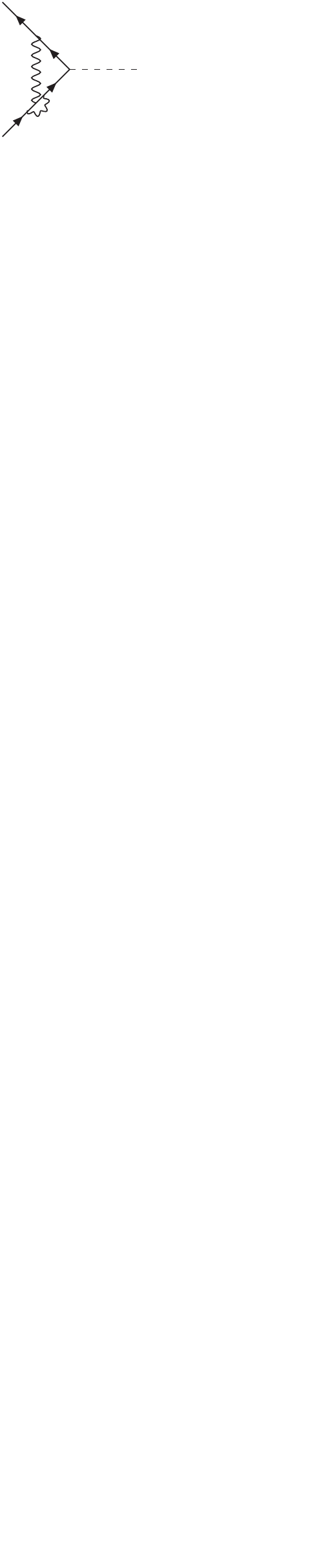}
\vspace*{-17cm}
  \caption{\sf The double virtual corrections of $O(a^2)$ to  $e^+e^-$ annihilation into a virtual 
gauge boson.The external self-energy corrections are not shown.}
  \label{fig:6}
\end{figure}

\begin{figure}[H]
  \centering
  \hskip-0.8cm
  \includegraphics[width=.24\linewidth]{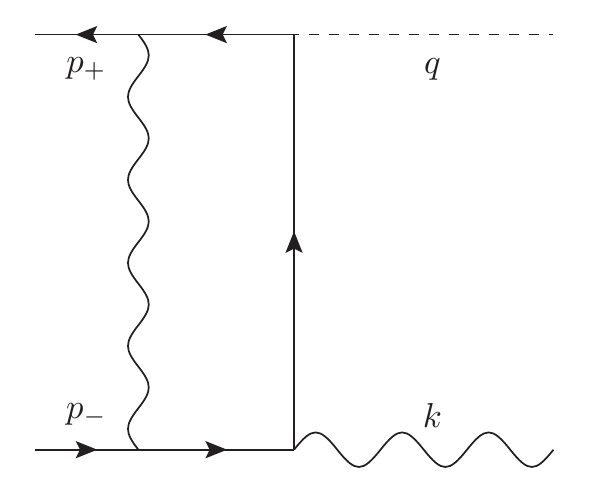}
  \includegraphics[width=.24\linewidth]{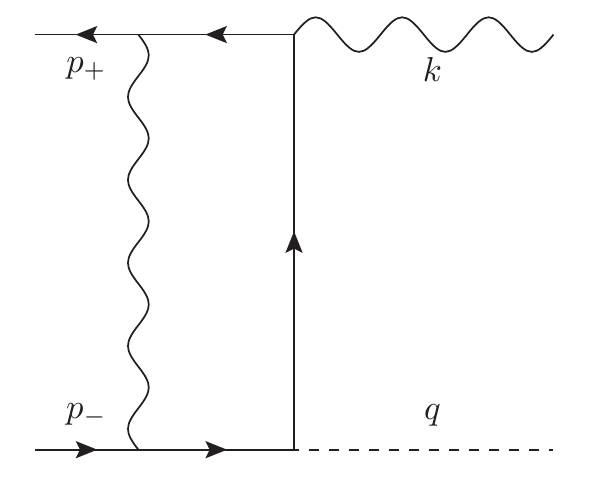}
  \includegraphics[width=.24\linewidth]{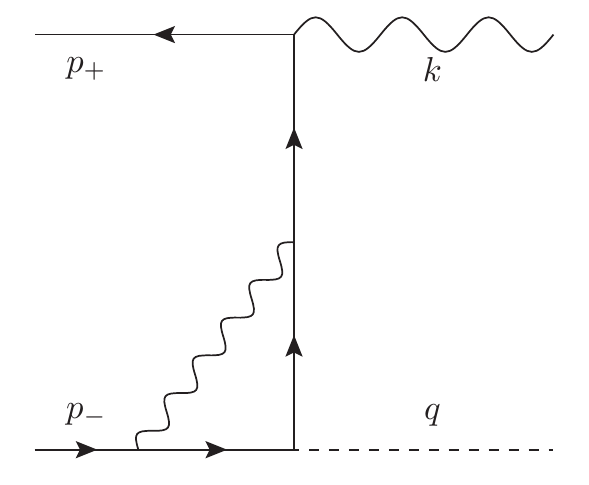}
  \includegraphics[width=.24\linewidth]{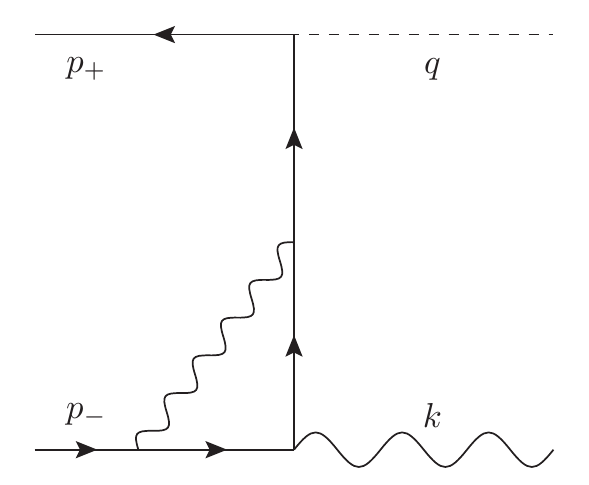} \hspace*{-10mm}
  \includegraphics[width=.24\linewidth]{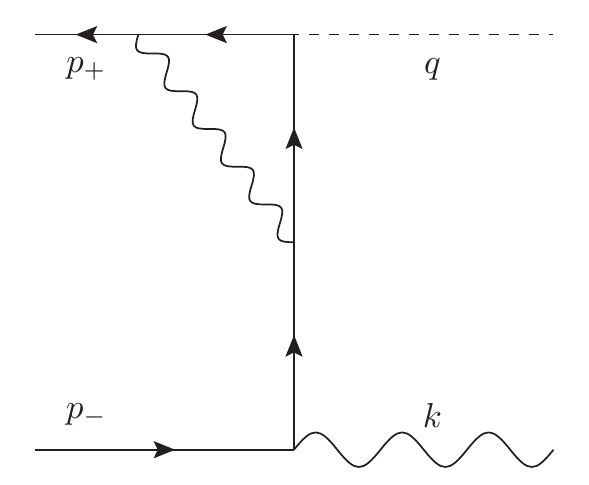}
  \includegraphics[width=.24\linewidth]{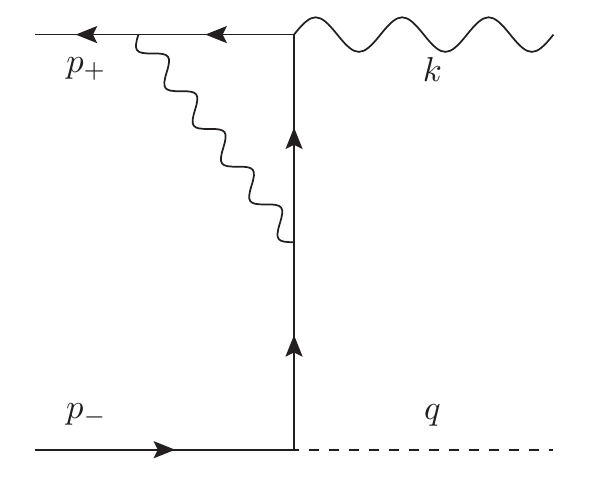}
  \includegraphics[width=.24\linewidth]{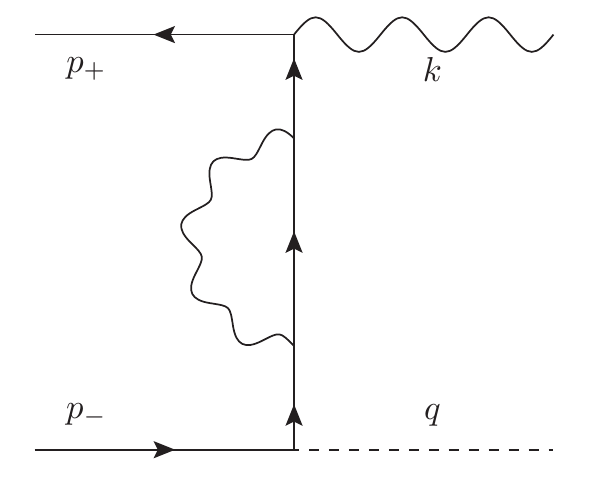}
  \includegraphics[width=.24\linewidth]{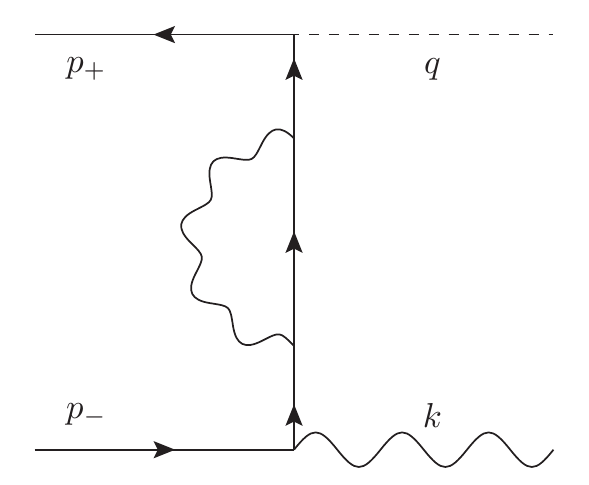}
  \caption{\sf The $O(a^2)$ virtual corrections to  $e^+e^-$ annihilation graphs into one photons and a virtual gauge 
boson.The external self-energy corrections are not shown.}
  \label{fig:7}
\end{figure}

\vspace*{1mm}
\noindent
The photonic two--loop corrections consist out of the following six contributions
\begin{eqnarray}
R_2^{\gamma\gamma} &=& 
T_2^{\rm S_2} +  
T_2^{\rm V_2} +
T_2^{\rm S_1 V_1} +
T_2^{\rm S_1 H_1} +
T_2^{\rm V_1 H_1} +
T_2^{\rm H_2},
\end{eqnarray}
with
\begin{enumerate}
\item $T_2^{\rm S_2}$:~~~~both emitted photons are soft,~~Figure~\ref{fig:gaga}
\item $T_2^{\rm V_2}$:~~~~both photons are virtual,~~Figure~\ref{fig:6}
\item $T_2^{\rm S_1 V_1}$:~~one photon is soft, one is virtual,~~Figure~\ref{fig:7}
\item $T_2^{\rm S_1 H_1}$:~~one photon is soft, one is hard,~~Figure~\ref{fig:gaga}
\item $T_2^{\rm V_1 H_1}$:~~one photon is virtual, one is hard,~~Figure~\ref{fig:7}
\item $T_2^{\rm H_2}$:\hspace*{.5mm}~~~~both emitted photons are hard,~~Figure~\ref{fig:gaga}.
\end{enumerate}
We will calculate the different contributions in this order. 

Due to the energy cut--off $\varepsilon$ on the photon energy all contributions
are functions. 
However, one can introduce $+$-distributions and drop the dependence
on the photon cut--off.
We still use the abbreviation
\begin{eqnarray}
{\cal D}_n(z) = \left(\frac{\ln^n(1-z)}{1-z}\right)_+,~~~n \in \mathbb{N},
\end{eqnarray}
besides of the $\delta(1-z)$-distribution.
However, using the $\varepsilon$ cut--off the $+$-distributions can be
understood as simple functions.

The double soft photon correction is obtained calculating the graphs given in Figure~\ref{fig:gaga}
in the soft photon approximation.
Since the soft corrections are factorizing, one obtains
\begin{eqnarray}
T_2^{\rm S_2} = \frac{1}{2} (T_1^{\rm S_1})^2 - 32 (L-1)^2 \zeta_2,
\end{eqnarray}
cf.~\cite{Berends:1987ab,Kuraev:1985hb,Jadach:1987ii}. Here the last term stems from the integral
\begin{eqnarray}
c_2 = - \lim_{\ep \rightarrow 0} 32 (L-1)^2 \int_\ep^\Delta \frac{dz_1}{z_1} \int_{\Delta-z_1}^\Delta \frac{dz_2}{z_2}
= -32 (L-1)^2 \zeta_2.
\end{eqnarray}

The diagrams for the double virtual corrections are shown in Figure~\ref{fig:6}, and
$T_2^{\rm V_2}$  is given by
\begin{eqnarray}
T_2^{\rm V_2} &=& |F_1(s,m_e,\lambda)|^2 + 2 {\sf Re} F_2(s,m_e,\lambda),
\end{eqnarray}
cf.~(\ref{eq:F1}) with
\begin{eqnarray}
{\sf Re} F_2(s,m_e,\lambda) &=& 16 \Biggl[ \frac{1}{32} L^4 - \frac{3}{16} L^3 + \left(\frac{17}{32} - \frac{5}{4} \zeta_2\right) 
L^2 + \Biggl( -\frac{21}{32} + 3 \zeta_2 + \frac{3}{2} \zeta_3\Biggr) L + \frac{2}{5} \zeta_2^2 - \frac{9}{4} \zeta_3 
\nonumber\\ &&
-3 \zeta_2 \ln(2) - \frac{1}{2} \zeta_2 + \frac{405}{216} + \frac{1}{8} \ln^2\left(\frac{\lambda^2}{m_e^2}\right)(L^2 - 2 L +1 -6 
\zeta_2)
- \frac{1}{8} \ln\left(\frac{\lambda^2}{m_e^2}\right)
\nonumber\\ && \times
(L^3 - 4 L^2 + (7-20 \zeta_2) L - 4 + 26 \zeta_2) \Biggr],
\end{eqnarray}
cf.~\cite{Barbieri:1972hn,Barbieri:1972as,Burgers:1985qg,Kniehl:1989kz}. Again, the Pauli Form Factor does not contribute
in the limit $s \gg m^2$. The soft corrections given in \cite{Burgers:1985qg} were corrected in \cite{Kniehl:1989kz}.
The virtual--soft corrections, Figure~\ref{fig:7}, are given by
\begin{eqnarray}
T_2^{\rm V_1 S_1} &=& T_1^{\rm S_1}~T_1^{\rm V_1}.
\end{eqnarray}

The virtual--hard corrections, Figure~\ref{fig:7}, read
\begin{eqnarray}
T_2^{\rm V_1 H_1} &=& 
 \frac{16 \big(1+z^2\big)}{1-z} (L-1)^2 \ln \left(\frac{\lambda ^2}{m_e^2}\right)
-\frac{8 \big(1+z^2\big)}{1-z} L^3
+L^2 \biggl(
         \frac{32 \big(1+z^2\big)}{1-z}
\nonumber \\ &&
        -\frac{8 \big(1+z^2\big)}{1-z}  \ln(z)
\biggr)
+L \biggl(
        -\frac{8 \big(7-z+8 z^2\big)}{1-z}
        +\frac{8 \big(3+4 z+3 z^2\big)}{1-z} \ln(z)
\nonumber \\ &&
        +\frac{16 \big(1+z^2\big)}{1-z}  \ln(1-z) \ln(z)
        -\frac{8 \big(1+z^2\big)}{1-z}  \ln^2(z)
        +\frac{16 \big(1+z^2\big)}{1-z} \text{Li}_ 2(1-z)
\nonumber \\ &&
        +\frac{64 \big(1+z^2\big)}{1-z}  \zeta_2
\biggr)
+\frac{8 \big(1+3 z+4 z^2\big)}{1-z}
-\frac{8 \big(1+12 z+3 z^2\big)}{1-z} \ln(z)
-\frac{8 \big(1+z^2\big)}{3 (1-z)} \ln^3(z)
\nonumber \\ &&
+\frac{32 z^2}{1-z} S_{1,2}(1-z)
-\frac{64 \big(1-z+2 z^2\big)}{1-z} \zeta_2
+\biggl(
         16 (1-z)
        -\frac{8 \big(2+6 z-3 z^2\big)}{1-z}  \ln(z)
\nonumber \\ &&
        +\frac{8 \big(1+z^2\big)}{1-z}  \ln^2(z)
\biggr) \ln(1-z)
+16 z \ln^2(1-z)
+\frac{4 \big(3+z^2\big)}{1-z}  \ln^2(z)
\nonumber \\ &&
+\biggl(
        -\frac{8 \big(2+6 z-3 z^2\big)}{1-z}
        -16 (1+z) \ln(1-z)
        +\frac{16 \big(1+z^2\big)}{1-z} \ln(z)
\biggr) \text{Li}_ 2(1-z)
\nonumber \\ &&
+32 (1+z) \text{Li}_ 3(1-z)
.
\end{eqnarray}

The two hard photon corrections, Figure~\ref{fig:gaga}, yield
\begin{eqnarray}
T_2^{\rm H_2} &=& 16 \Biggl\{ - 2 \frac{1+z^2}{1-z} (L-1)^2 \ln(\ep)
+ \Biggl\{\frac{1+z^2}{1-z} \left[2 \ln(1-z) -\frac{1}{2} \ln(z)\right] - 1+z +\frac{1}{2}(1+z) 
\nonumber\\ &&
\times \ln(z) \Biggr\} L^2
+\Biggl\{ \frac{1+z^2}{1-z}[-4\ln(1-z) +\ln(z)] - z\ln(z) +\frac{7}{2}(1-z) + \frac{1}{4}(1+z) 
\nonumber\\ &&
\times \ln^2(z) \Biggr\} L
- \frac{1+z^2}{1-z} \Biggl[S_{1,2}(1-z) + \frac{1}{2} \ln(z) \Li_2(1-z) + \frac{3}{2} \ln^2(z) - \left(\frac{7}{6} + 
\zeta_2\right) 
\nonumber\\ &&
\times \ln(z) 
- 2 \ln(1-z) \Biggr] 
- \frac{1}{2}(1+z) (\Li_3(1-z) + 2 S_{1,2}(1-z)) 
-\frac{2}{3} z\zeta_2 
\nonumber\\ &&
- \frac{1}{6} (3+4z) \Li_2(1-z) 
- \frac{1}{6} (10 - 25 z) \ln(z)  
+ \frac{2}{(1-z)^2} \ln^2(z) - \frac{1}{12}(3 + 28 z) \ln^2(z) 
\nonumber\\ &&
+ \frac{1}{2}(1-z) - \frac{2}{3} \frac{z}{1-z} \Biggl[1 + \frac{2}{1-z} \ln(z) 
+ \frac{1}{(1-z)^2} \ln^2(z) \Biggr] \Biggr\} \nonumber\\
&=& 64 {\cal D}_1(z) (L-1)^2 + \tilde{T}_2^{\rm H_2}.
\end{eqnarray}
All corrections but the virtual--hard corrections agree with the results in \cite{Berends:1987ab}.

The complete photonic corrections are given by
\begin{eqnarray}
R_2^{\gamma\gamma}&=&
                  \delta(1-z) \Bigl\{
                        32 (L-1)^2 \ln^2(\varepsilon)
                        + \big( 
                              48 L^2 - (112-64\zeta_2) L + 64 - 64 \zeta_2
                        \big) \ln(\varepsilon)
\nonumber \\ &&
                        + (18-32\zeta_2) L^2
                        - (45-88\zeta_2-48\zeta_3) L
                        + 76 + (6-96\ln(2))\zeta_2 -72 \zeta_3 - \frac{96}{5} \zeta_2^2
                  \Bigr\}
\nonumber \\ &&
                  + \theta(1-z-\varepsilon) \Bigl\{
                        64 (L-1)^2 {\cal D}_1
                        + \big(
                              48 L^2 - (112-64\zeta_2) L + 64 - 64 \zeta_2
                        \big) {\cal D}_0
\nonumber \\ &&
                        - L^2 \biggl(
                               8 (5+z)
                              +32 (1+z) \ln (1-z)
                              +\frac{8 \big(1+3 z^2\big)}{1-z} \ln(z)
                        \biggr)
                        + L \biggl(
                              8 (14+z)
\nonumber \\ &&
                              +\frac{8 \big(5+2 z+7 z^2\big)}{1-z}  \ln(z)
                              -\frac{4 \big(1+3 z^2\big)}{1-z}  \ln^2(z)
                              +\frac{16 \big(1+z^2\big)}{1-z} \text{Li}_2(1-z)
\nonumber \\ &&
                              -32 (1+z) \zeta_2
                              +\biggl[
                                    64 (1+z)
                                    +\frac{16 \big(1+z^2\big)}{1-z}  \ln(z)
                              \biggr] \ln(1-z)
                        \biggr)
                        -\frac{8 \big(18+z-15 z^2\big)}{3 (1-z)}
\nonumber \\ &&
                        -\frac{8 \big(1+z^2\big)}{3 (1-z)}  \ln^3(z)
                        +\frac{4}{3 (1-z)^3} \big(12-33 z+51 z^2-51 z^3+13 z^4\big) \ln^2(z)
\nonumber \\ &&
                        +\frac{32}{3} (3+8 z) \zeta_2
                        -32 (1+z) \zeta_3
                        -\biggl(
                               16 (1+3 z)
                              +\frac{8 \big(2+6 z-3 z^2\big) }{1-z} \ln(z)
\nonumber \\ &&
                              -\frac{8 \big(3-z^2\big)}{1-z}  \ln^2(z)
                        \biggr) \ln(1-z)
                        +16 z \ln^2(1-z)
                        -\biggl(
                               \frac{8 \big(6+3 z+26 z^2-27 z^3\big)}{3 (1-z)^2}
\nonumber \\ &&
                              +\frac{16 \big(1-3 z^2\big)}{1-z} \zeta_2
                        \biggr) \ln(z)
                        +\biggl(
                              -\frac{8 \big(9+19 z-13 z^2\big)}{3 (1-z)}
                              -16 (1+z) \ln(1-z)
\nonumber \\ &&
                              +\frac{8 \big(5-3 z^2\big)}{1-z} \ln(z)
                        \biggr) \text{Li}_2(1-z)
                        +24 (1+z) \text{Li}_3(1-z)
                        +32 (1+z) \text{Li}_3(z)
                  \Bigr\}.
\end{eqnarray}
We now turn to the fermion--pair emission contributions in different channels.

\section{\boldmath $O(\alpha^2)$ Non--Singlet Corrections due to $e^+e^-$ Emission}
\label{sec:6}
\begin{figure}[H]
  \centering
  \hskip-0.8cm
  \vspace*{5mm}  
  \begin{minipage}[h]{0.8\textwidth} 
  \includegraphics[width=\linewidth]{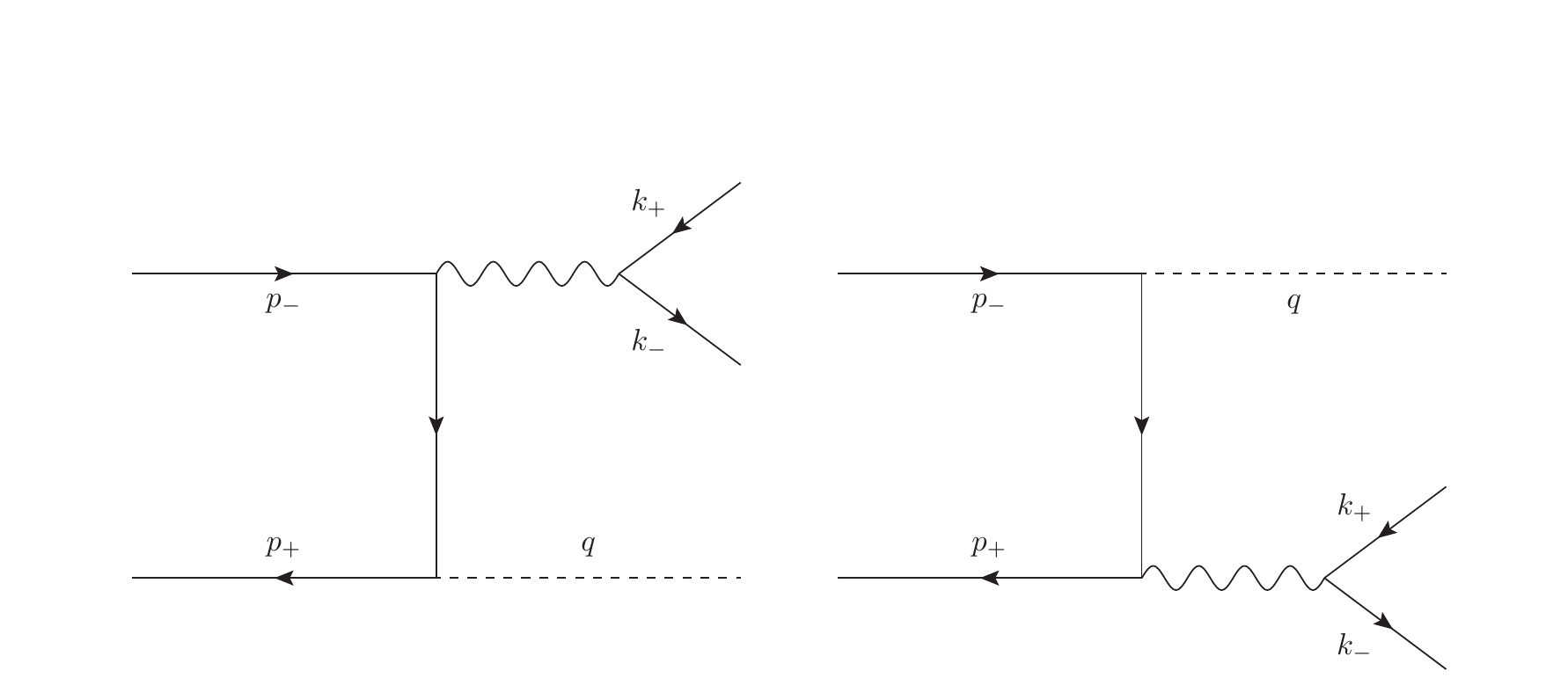} 
  \end{minipage}
  \caption{\sf The $e^+e^-$ annihilation graphs into a fermion pair and a gauge boson (process $A$).}
  \label{fig:A}
\end{figure}

\vspace*{1mm}
\noindent
The non--singlet contributions can be given by
\begin{eqnarray}
  R_2^{e^+e^-,\rm NS} = \delta(1-z) \left( R_2^{e^+e^-,\rm NS,\rm S} + R_2^{e^+e^-,\rm NS, \rm V} \right)
  + \theta(1-z-\varepsilon) R_2^{e^+e^-,\rm NS,\rm H},
\end{eqnarray}
where $R_2^{e^+e^-,\rm NS,\rm S}$, $R_2^{e^+e^-,\rm NS,\rm V}$ and $R_2^{e^+e^-,\rm NS,\rm H}$
denote corrections due to soft, virtual and hard fermion-pair radiation respectively.
The first two contributions were correctly given in \cite{Berends:1987ab} and read
\begin{eqnarray}
  R_2^{e^+e^-,\rm NS,\rm S} &=& \frac{8}{9} L^3 - \frac{40}{9} L^2 + \biggl( \frac{448}{27} - \frac{32}{3} \zeta_2 \biggr) L
  - \frac{2624}{81} + \frac{160}{9} \zeta_2 + \frac{64}{3} \zeta_3 + \frac{64}{9} \ln^3(\varepsilon) 
  \nonumber \\ &&
  - \biggl( \frac{160}{9} - \frac{32}{3} L \biggr) \ln^2(\varepsilon)
  + \biggl( \frac{896}{27} - \frac{160}{9} L + \frac{16}{3} L^2 - \frac{64}{3} \zeta_2 \biggr) \ln(\varepsilon) 
\\
  R_2^{e^+e^-,\rm NS,\rm V} &=& - \frac{8}{9} L^3 + \frac{76}{9} L^2 - \frac{4}{27} \left( 265-72\zeta_2\right)L+\frac{3064}{27} - \frac{176}{3}\zeta_2
\end{eqnarray}
In the following we will calculate the hard contributions.
They can be expressed by iterative integrals $\HA_{\vec{a}}(u) \equiv \HA_{\vec{a}}$ 
up to weight {\sf w = 2} over the alphabet given in Section~\ref{sec:4} and $u = 4 \rho/(1 - \sqrt{z})^2$. The 
contributing diagrams are shown in Figure~\ref{fig:A}.

The correction to the scattering cross section for electron pair emission is given by
\begin{align}
  R_2^{e^+e^-,\rm NS,\rm H} &=
        \int\limits_{4 m_f^2}^{s(1-\sqrt{z})^2} d \spp
        \frac{16}{3 s \left. \spp \right. ^2}
        \sqrt{ 1 - \frac{4 m_f^2}{\spp} } ( 2 m_f^2 + \spp )
        \Biggl\{
\nonumber \\
&\hspace{-1cm}- \frac{ \lambda^{1/2}(s,\sp,\spp) \bigl[ 2 s \, \sp \, \spp + m_i^2 \bigl( s^2 + (\sp-\spp)^2 \bigr) + 4 s \, 
m_i^4 \bigr] }{ s \, \sp \, \spp + m_i^2 \bigl( s^2 + (\sp-\spp)^2 - 2 s \, (\sp+\spp) \bigr) }
\nonumber \\
&\hspace{-1cm}+ \frac{ (\sp+\spp)^2 + 4 m_i^2 (s-\sp-\spp)+s^2-8 m_i^4 }{\beta(s-\sp-\spp)} 
\ln\left( \frac{s-\sp-\spp+\beta 
\lambda^{1/2}(s,\sp,\spp)}{s-\sp-\spp-\beta \lambda^{1/2}(s,\sp,\spp)} \right)
\Biggr\},
\label{eq:DY-NS-Integrand}
\end{align}
with
\begin{eqnarray}
\beta &=& \sqrt{ 1 - \frac{4m_e^2}{s} }
\end{eqnarray}
and
\begin{eqnarray}
\lambda(s,\sp,\spp) &=& s^2+\left.\sp\right.^2+\left.\spp\right.^2-2s\sp-2s\spp-2\sp\spp.
\end{eqnarray}

One performs the transformation
\begin{eqnarray}
t &=& \frac{4m^2}{\spp},~~~~~z =\frac{\sp}{s} 
\end{eqnarray}
in order to introduce dimensionless quantities. This yields
\begin{eqnarray}
\label{eq:NS1}
R_2^{e^+e^-,\rm NS,\rm H} &=& 
\int\limits_{t_-}^{t^+} dt \frac{8}{3 t^2} \sqrt{1-t} (2+t) \Biggl\{
\frac{ 16 \rho^2 + 4 \rho t ( t -2 z) + t \bigl( t + z (8 + z t) \bigr) }{ 16 \rho^2 - 8 \rho (1 + z) t + t \bigl(4 z + (1 - z)^2 
t \bigr)} 
\nonumber\\ && \times
\sqrt{16 \rho^2-8 \rho (z+1) t+(z-1)^2 t^2}
\nonumber \\
&& - \frac{8 \rho^2 \bigl( t (t+2) - 2 \bigr) + 4 \rho t \bigl( z (t-2)-t \bigr) - \left(z^2+1\right) t^2}{\sqrt{1-4 \rho} (4 
\rho - ( 1 - z) t) }
\nonumber \\
&& \times \ln\left( \frac{4 \rho-(1-z) t -\sqrt{1-4 \rho} \sqrt{16 \rho^2-8 \rho (1+z) t + (1-z)^2 t^2} }{4 \rho - (1-z)t + 
\sqrt{1-4 \rho} 
\sqrt{16 
\rho^2-8 \rho (1+z) t+(1-z)^2 t^2}} \right)
\Biggr\},
\end{eqnarray}
with
\begin{eqnarray}
t_- &=& 1,~~~~~t^+ = \frac{4 \rho}{(1-\sqrt{z})^2}.
\end{eqnarray}
Integrating this expression exactly, one obtains
\begin{align}
        R_2^{e^+e^-,\rm NS} &=
        \Biggl\{
         \frac{64}{3} z (1-z) (1+z-4 \rho) \HT_{v_4,d_7} 
        +\frac{256}{3} z \rho  (1 + z - 4 \rho) \HT_{v_4,d_6} 
\nonumber \\ &
        +\frac{128 z (1 - 4 \rho^2 ) (1-z+2 \rho) (1-z-4 \rho)}{3 (1-z)^2} \HT_{d_8,d_7} 
\nonumber \\ &
        +\frac{512 z \rho  ( 1 - 4 \rho^2 ) (1-z+2 \rho) (1-z-4 \rho)}{3(1-z)^3} \HT_{d_8,d_6} 
\nonumber \\ &
        +\frac{16}{9 (1-z)^2}
        \Bigl[
                (1+z)^2 \big(4-9 z+4 z^2\big) +2 \big(9-16 z+13 z^2-2 z^3\big) \rho +32 \rho ^2
        \Bigr] \HT_{d_2} 
\nonumber \\ &
        +\frac{512 z \rho}{9 (1-z)^4}
        \Bigl[
                3 (1-z)^4 z
                - (1-z)^3 \big(4+z^2\big) \rho
                -2 \big(9-29 z+38 z^2-17 z^3+3 z^4\big) \rho^2
\nonumber \\ &
                -4 (2-z) \big(3+6 z-5 z^2\big) \rho^3
                +16 \big(7-8 z+9 z^2\big) \rho^4
                +128 (3-z) \rho^5
        \Bigr] \HT_{d_4} 
\nonumber \\ &
        -\frac{16}{9 (1-z)^4}
        \Bigl[
                3-34 z+129 z^2-212 z^3+129 z^4-34 z^5+3 z^6
                + 8 \big(2-16 z+9 z^2
\nonumber \\ &
                +4 z^3-5 z^4+2 z^5\big) \rho
                +16 z \big(12-13 z+18 z^2-z^3\big) \rho ^2
                +32 \big(1+22 z-7 z^2\big) \rho ^3
        \Bigr] \HT_{d_1} 
\nonumber \\ &
        -\frac{128 z}{9 (1-z)^4}
        \Bigl[
                1+7 z-47 z^2+86 z^3-47 z^4+7 z^5+z^6
                - 2 \big(7-55 z+54 z^2
\nonumber \\ &
                +16 z^3-17 z^4+3 z^5\big) \rho
                - 4 \big(39-16 z+16 z^2+4 z^3+5 z^4\big) \rho ^2
\nonumber \\ &
                +16 \big(8-23 z+22 z^2+9 z^3\big) \rho ^3
                +128 \big(7+2 z-z^2\big) \rho ^4
        \Bigr] \HT_{d_5} 
        -\frac{64}{3} (2 z+(1-z) \rho) \HT_{d_3} 
\nonumber \\ &
        + \biggl[ \frac{16}{3 \sqrt{1-4 \rho }} (1+z-4 \rho) \HT_{v_4} 
                + \frac{32 (1-4 \rho^2 ) (1-z+2 \rho) (1-z-4 \rho)}{3 (1-z)^3 \sqrt{1-4 \rho }} \HT_{d_8} 
        \biggr]
\nonumber \\ & \times
        \ln \left( \frac{1-z-4 \rho-\sqrt{1-4 \rho } \sqrt{(1-z)^2-8 (1+z) \rho+16 \rho ^2}}{1-z-4 \rho+\sqrt{1-4 \rho } 
\sqrt{(1-z)^2-8 (1+z) \rho+16 \rho ^2}} \right)
        \Biggr\}.
\label{eq:DY-NS-FULL}
\end{align}
To obtain the expansion in the case of the emission of an $e^+e^-$ pair one cannot simply set $m_e \rightarrow 0$ in (\ref{eq:NS1}) as 
has
been done in Refs.~\cite{Berends:1987ab,Kniehl:1988id}; see, however, Section~\ref{sec:7}. 
One first rewrites the integral (\ref{eq:NS1}) by
\begin{eqnarray}
\label{eq:NS2}
\int_{t_-}^{t_+} dt [f(t,\rho,z) - f_\rho(t,\rho,z)] + \int_{t_-}^{t_+} dt f_\rho(t,\rho,z).
\end{eqnarray}
Here $f_\rho(t,\rho,z)$ denotes the integrand $f(t,\rho,z)$, expanded in $\rho$, including the $\rho^0$ term.
For the first term in Eq.~(\ref{eq:NS2}) the variable transformation
\begin{eqnarray}
\label{eq:NS2}
t = \frac{1}{1+(\xi^{-1}-1) \bar{t}},~~~~\xi = \frac{4\rho}{(1-\sqrt{z})^2}
\end{eqnarray}
is performed leading to $\bar{t} \in [0,1]$. After that, the integrand can be expanded in $\rho$.
A further variable transformation is necessary to rationalize the root
\begin{eqnarray}
\label{eq:NS3}
\sqrt{(\bar{t}-1)[\bar{t}(1-\sqrt{z})^2 - (1+\sqrt{z})^2]}.
\end{eqnarray}
One may choose 
\begin{eqnarray}
\label{eq:NS4}
\bar{t} = \frac{(1-v)(v-z)}{v(1-\sqrt{z})^2},~~~~v \in [z, \sqrt{z}].
\end{eqnarray}
The integral can now be performed. 
This leads to the correct result in the limit $m_e^2 \ll s$.
Of course we can also expand Eq.~(\ref{eq:DY-NS-FULL}) in $\rho$.
We checked that both methods agree.
Including the term $\rho^0$ one obtains
\begin{align}
\label{eq:R2ee1}
R_2^{e^+e^-,\rm NS, \rm H}
&=
 \frac{8 \big(1+x^2\big)}{3 (1-x)} L^2
 +\biggl[
        -\frac{16 \big(11-12 x+11 x^2\big)}{9 (1-x)}
        +\frac{32 \big(1+x^2\big)}{3 (1-x)} \ln(1-x)
\nonumber \\ &
        -\frac{16 \big(1+x^2\big)}{3 (1-x)}  \ln(x)
\biggr] L
+\frac{32 \big(1+x^2\big)}{3 (1-x)}  \ln^2(1-x)
-\frac{16 x^2}{3 (1-x)}  \text{Li}_2(1-x)
\nonumber \\ &
+\frac{32}{9 (1-x)^3} \big(7-13 x+8 x^2-13 x^3+7 x^4\big)
-\frac{16 x}{9 (1-x)^4} \big(3-36 x+94 x^2
\nonumber \\ &
-72 x^3+19 x^4\big)  \ln(x)
-\frac{32 \big(1+x^2\big)}{3 (1-x)} \zeta_2
-\biggl(
        \frac{32 \big(11-12 x+11 x^2\big)}{9 (1-x)}
\nonumber \\ &
        +\frac{32 \big(1+x^2\big)}{3 (1-x)}  \ln(x)
\biggr) \ln(1-x)
-\frac{8 x^2}{3 (1-x)}  \ln^2(x)
        + \mathcal{O}\left( \rho L^2 \right) .
\end{align}
This result differs from the one presented in Refs.~\cite{Berends:1987ab,Kniehl:1988id} exactly by the term given 
in \cite{Blumlein:2019srk}, Eq.~(8) and agrees with the result obtained in Ref.~\cite{Blumlein:2011mi} based 
on massive operator matrix elements. The reason for this disagreement lays in the neglection of some of the electron 
mass terms before all integrals have been performed.
The full radiator therefore reads
\begin{eqnarray}
  R_2^{e^+e^-,\rm NS} &=& \delta(1-z) \Biggl(
    4 L^2 - \frac{68}{3} L + \frac{6568}{81} - \frac{368}{9} \zeta_2
    + \frac{64}{3} \zeta_3 + \frac{64}{9} \ln^3(\varepsilon)
    + \biggl( \frac{32}{3} L - \frac{160}{9} \biggr) \ln^2(\varepsilon)
    \nonumber \\ &&
    + \biggl( \frac{16}{3} L^2 - \frac{160}{9} L + \frac{896}{27} - \frac{64}{3} \zeta_2\biggr) \ln(\varepsilon)
    \Biggr) + \theta(1-z-\varepsilon) R_2^{e^+e^-,\rm NS, \rm H}
\end{eqnarray}
\section{\boldmath Heavier Fermionic Final States in the Non--Singlet Process}
\label{sec:7}
  
\vspace*{1mm}  
\noindent   
If in $e^+e^-$ annihilation a heavier fermion pair $f\bar{f}$ with $m_f \gg m_e$ is radiated via a virtual 
photon from the initial state 
electrons
 one may use, cf.~\cite{Kniehl:1988id} and \cite{Berends:1987ab}, Erratum,
\begin{eqnarray}
\frac{d{\sigma}^{\rm II}}{d\sp} &=& 
\hspace*{-2mm}
a^2 \frac{\sigma_0{s'}}{s}
N_{C,f} Q_f^2 \int\limits_{4m^2}^{s(1-\sqrt{z})^2} d \spp \frac{16}{3 s \left. 
\spp \right. ^2} \sqrt{ 1 - \frac{4m^2}{\spp} } ( 2m^2 + \spp ) 
\nonumber \\
&& \hspace*{-2mm}
\times \Biggl\{ 
- 2 \lambda^{1/2}(s,\sp,\spp) + \frac{ s^2 + (\sp + \spp)^2 }{s-\sp-\spp} \ln\left( 
\frac{s-\sp-\spp+\lambda^{1/2}(s,\sp,\spp)}{s-\sp-\spp-\lambda^{1/2}(s,\sp,\spp)} \right)
\Biggr\},~~~m \gg m_e. \nonumber \\
&=& \hspace*{-2mm}
N_{C,f} Q_f^2 \int\limits_{t_-}^{t^+} dt \frac{8}{3 t^2} \sqrt{1-t} (2+t) \Biggl\{ 
2 \sqrt{16 \rho^2-8 \rho (1+z) t+(1-z)^2 t^2}
\nonumber \\
&&\hspace*{-2mm}
+ \frac{16 \rho^2+8 \rho z t+\left(z^2+1\right) t^2}{4 \rho-(1-z) t}
\ln\left( \frac{4 \rho-(1-z) t - \sqrt{16 \rho^2-8 \rho (1+z) t + (1-z)^2 t^2} }{4 \rho - (1-z)t + \sqrt{16 \rho^2-8 r (1+z) 
t+(1-z)^2 
t^2}} \right)
\Biggr\},
\nonumber\\
\end{eqnarray}
with $N_{C,f} = 1, Q_f = -1$ for $f = \mu, \tau$ and $N_{C,f} = 3, Q_f = (2/3, -1/3, 2/3)$ for $f = c,b,t$.
Here we consider the case of heavier charged lepton pairs and heavy quark $(c,b,t)$ pairs. One obtains 
\begin{align}
\label{eq:BK1}
        \frac{d \sigma^{(2),\text{II}} (z,0,m_f=m) }{d \sp}
        &=
        a^2 \frac{\sigma^{(0)}(s^\prime)}{s}  \, \int\limits_{4 m^2}^{s(1-\sqrt{z})^2} d \spp
        \frac{16}{3 s \left. \spp \right. ^2}
        \sqrt{ 1 - \frac{4m^2}{\spp} } ( 2m^2 + \spp )
        \Biggl\{
\nonumber \\   
        &- 2 \lambda^{1/2}(s,\sp,\spp) + \frac{ s^2 + (\sp + \spp)^2 }{s-\sp-\spp} \ln\left( 
\frac{s-\sp-\spp+\lambda^{1/2}(s,\sp,\spp)}{s-\sp-\spp-\lambda^{1/2}(s,\sp,\spp)} \right)
\Biggr\}.
\end{align}
This result agrees with those of Refs.~\cite{Berends:1987ab,Kniehl:1988id}.

One may derive it also using the method of massive operator matrix elements. Here, the external lines in Figure~7, 
\cite{Blumlein:2011mi}, have to be taken massless, since $m_e \ll m$, and $m$ denotes the mass of the internal fermion line.
$\Gamma_{ee}^{(1),II}$ in (41), \cite{Blumlein:2011mi}, reads then \cite{Bierenbaum:2009mv}
\begin{eqnarray}
A_{ee,\mu}^{\overline{\rm MS},\rm II} 
= \frac{\beta_{0,H}}{4} \gamma_{ee}^{(0)}(N) \ln^2\left(\frac{m_\mu^2}{\mu^2}\right)
+ \frac{1}{2} \hat{\gamma}_{ee}^{(1),\rm II}(N) \ln\left(\frac{m_\mu^2}{\mu^2}\right) + a_{ee,\mu}^{(2),\rm NS} 
- \frac{\beta_{0,H}}{4} 
\gamma_{ee}^{(0)}(N) \zeta_2
\end{eqnarray}
and Eq.~(75) in \cite{Blumlein:2011mi} has to be replaced by Eq.~(4.16) in  \cite{Bierenbaum:2009mv} for QED.
Here the relation
\begin{eqnarray}
P_{ij}^{(k)}(N) = - \gamma_{ij}^{(k)}(N) 
\end{eqnarray}
holds and
\begin{eqnarray}
\hat{\gamma}_{ij}^{(k)} = 
{\gamma}_{ij}^{(k)}(N_F+1) - {\gamma}_{ij}^{(k)}(N_F), 
\end{eqnarray}
where $N_F$ denotes the number of massless fermions, which is $N_F = 1$ here.
The term $T_{\rm II}^{\mu^+\mu^-}$ corresponding to the one in (\ref{eq:R2ee1}) for $\mu^+\mu^-$ pair radiation reads
\begin{eqnarray}
T_{\rm II}^{\mu^+\mu^-} &=& 16\Biggl\{\frac{1}{6} \frac{1+z^2}{1-z} L^2_\mu 
+ \Biggl[\frac{1}{3} \frac{1+z^2}{1-z} \left(2 \ln(1-z) - \ln(z) - \frac{5}{3}\right) - \frac{2}{3}(1-z)\Biggr] L_\mu
\nonumber\\ &&
\frac{1+z^2}{1-z} \Biggl[ 
\frac{2}{3} \ln^2(1-z) - \frac{2}{3} \ln(z) \ln(1-z) + \frac{1}{12} \ln^2(z) - \frac{10}{9} \ln(1-z)
+ \frac{5}{9} \ln(z) 
\nonumber\\ &&
- \frac{1}{6} \Li_2(1-z) - \frac{2}{3} \zeta_2 + \frac{28}{27} \Biggr] 
- \frac{1}{3} (1-z) \left(4 \ln(1-z) - 2 \ln(z)  - \frac{19}{3} \right) - \frac{1}{3} \ln(z) 
\nonumber\\ &&
+ \frac{1}{6} (1+z) 
\left(\frac{1}{2} \ln^2(z) + \Li_2(1-z)\right)\Biggr\},
\end{eqnarray}
with $L_\mu = \ln(s/m_\mu^2)$.
\section{\boldmath The pure singlet corrections}
\label{sec:8}

\begin{figure}[H]
  \centering
  \begin{minipage}[h]{0.7\textwidth} 
  \includegraphics[width=\linewidth]{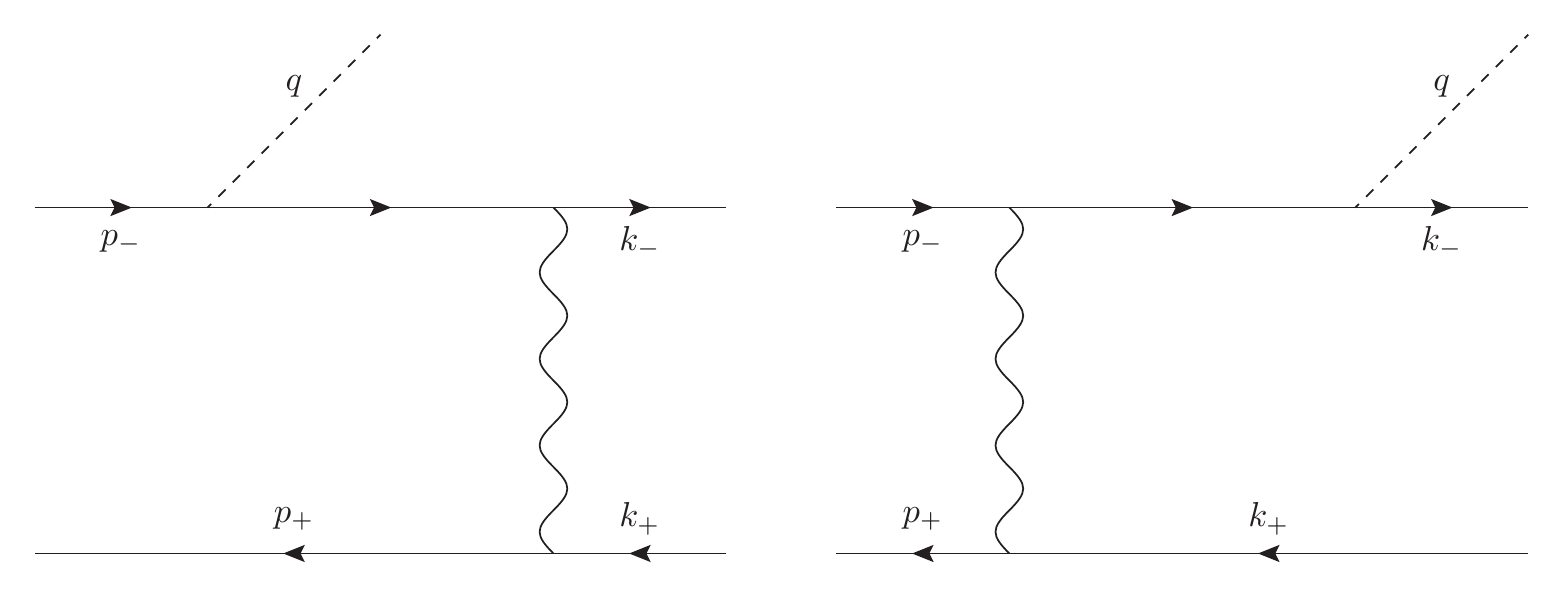} 
  \end{minipage}
\vspace*{5mm}
  \begin{minipage}[h]{0.7\textwidth} 
  \includegraphics[width=\linewidth]{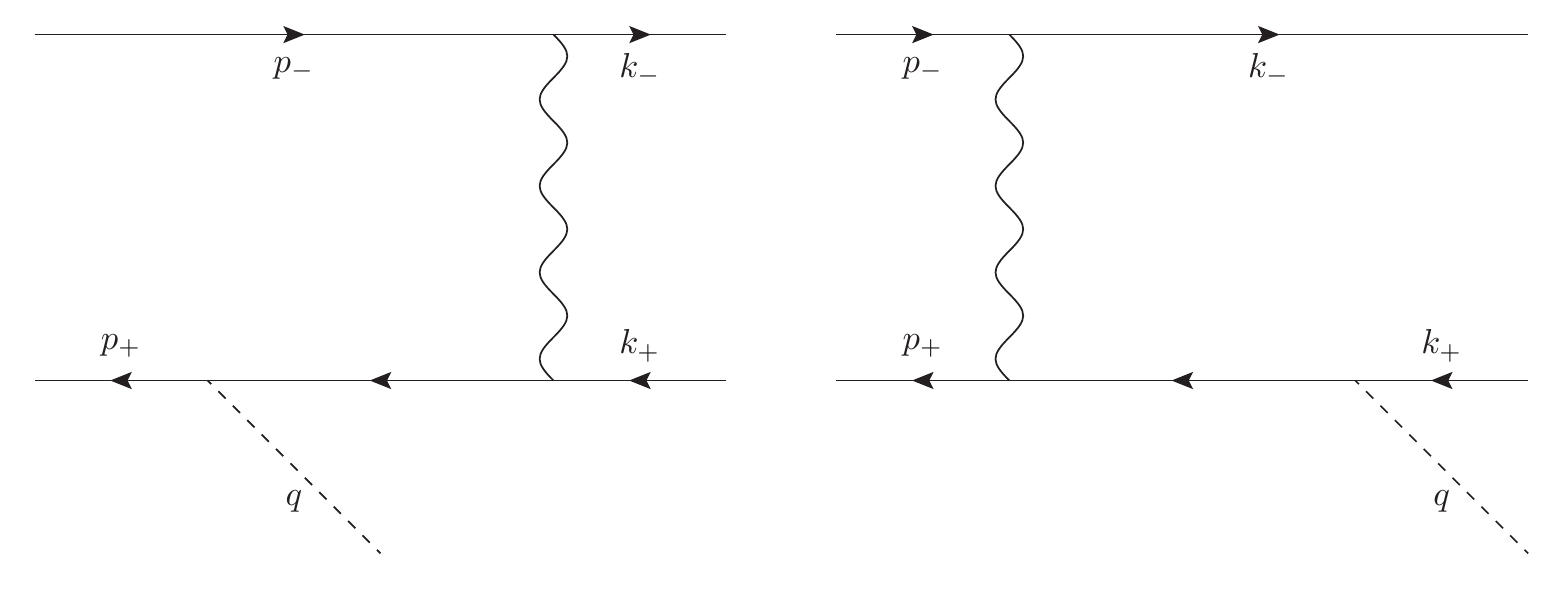} 
  \end{minipage}
  \caption{\sf The $e^+e^-$ annihilation graphs into a fermion pair and a gauge boson (process $A$).}
  \label{fig:B1}
\end{figure}

\vspace*{1mm}
\noindent
The diagrams contributing to the pure singlet case are shown in Figure~\ref{fig:B1}. Here one has to distinguish
between the vector and axial--vector case since different corrections are obtained. 
The radiator in the vector case is given by
\begin{eqnarray}
 R_2^{e^+ e^-,\rm v, \rm PS} &=&
       \biggl[
             \frac{4 (1-z) \big(4+7 z+4 z^2\big)}{3 z}
             \nonumber \\ &&
            +8 (1+z) \ln(z)
      \biggr] L^2
      -\biggl[
             \frac{128 (1-z) \big(1+4 z+z^2\big)}{9 z}
             \nonumber \\ &&
            +\frac{8 \big(4+6 z-3 z^2-8 z^3\big)}{3 z} \ln(z)
            -\biggl(
                   \frac{16 (1-z) \big(4+7 z+4 z^2\big)}{3 z}
                   \nonumber \\ &&
                  +32 (1+z) \ln (z)
            \biggr) \ln (1-z)
            +16 (1+z) \ln^2(z)
            \nonumber \\ &&
            -32 (1+z) \text{Li}_ 2(1-z)
      \biggr] L
      -\frac{4 \big(12+21 z-27 z^2-4 z^3\big)}{3 z}  \ln^2(z)
      \nonumber \\ &&
      +\frac{2 (1-z)}{27 z (1+z)^2} \big(-80+2463 z+5041 z^2+2949 z^3+163 z^4\big)
      \nonumber \\ &&
      -\frac{16 (1-z) \big(2-z+2 z^2\big)}{3 z} \zeta_2
      +96 (1+z) \zeta_3
      -\biggl(
             \frac{256 (1-z) \big(1+4 z+z^2\big)}{9 z}
             \nonumber \\ &&
            +\frac{16 \big(4+6 z-3 z^2-8 z^3\big)}{3 z} \ln(z)
            +32 (1+z) \ln^2(z)
      \biggr) \ln(1-z)
      \nonumber \\ &&
      +\biggl(
             \frac{16 (1-z) \big(4+7 z+4 z^2\big)}{3 z}
            +32 (1+z) \ln(z)
      \biggr) \ln^2(1-z)
      \nonumber \\ &&
      +\biggl(
             \frac{64 (1-z) \big(1+4 z+z^2\big)}{3 z} \ln(1+z)
            -\frac{4}{9 z (1+z)^3} \big(40+3 z-345 z^2
            \nonumber \\ &&
            -445 z^3+213 z^4+318 z^5+64 z^6\big)
      \biggr) \ln(z)
      -8 (1+z) \ln^3(z)
      \nonumber \\ &&
      +\biggl(
            \frac{8 \big(-4-9 z+3 z^2+12 z^3\big)}{3 z}
            +64 (1+z) \ln(1-z)
            \nonumber \\ &&
            -48 (1+z) \ln(z)
      \biggr) \text{Li}_ 2(1-z)
      +\biggl(
             \frac{64 (1-z) \big(1+4 z+z^2\big)}{3 z}
             \nonumber \\ &&
            -64 (1+z) \ln(z)
      \biggr) \text{Li}_ 2(-z)
      -64 (1+z) \text{Li}_ 3(1-z)
      \nonumber \\ &&
      +128 (1+z) \text{Li}_ 3(-z)
      -32 (1+z) S_{1,2}(1-z)
      + T_{\rm III, v, int}
\label{eq:PSv}
\end{eqnarray}
and consists of the direct terms and the interference terms
\begin{eqnarray}
T_{\rm III, v, int} &=&
\Biggl\{\Biggl(2 + z + \frac{2}{z}\Biggr) \Bigl[
32 S_{1,2}(1-z) - 96 S_{1,2}(-z) - 48 \ln^2(1+z) \ln(z) - 48 \zeta_2 \ln(1+z) 
\nonumber\\ &&
+ 40 \ln^2(z) \ln(1+z) - 96 \Li_2(-z) 
\ln(1+z) \Bigr]
+40(1+z)\Bigl[2 \Li_2(-z) + 2 \ln(z) \ln(1+z) 
\nonumber\\ &&
+\zeta_2\Bigr]
-8 \Biggl(6 -3z- \frac{4}{z}\Biggr) \Li_3(1-z) + 16\Biggl(10 - 3z - \frac{10}{z}\Biggr) \Li_3(-z)
+ 24 \Biggl(6 - z - \frac{4}{z} \Biggr) \zeta_3 
\nonumber\\ &&
+ 8(10-z) \Li_2(1-z) \ln(z) -\frac{16}{3} z \ln^3(z)
+ 32 \Biggl( 2z + \frac{5}{z}\Biggr) \Li_2(-z) \ln(z) 
-52 z \ln^2(z) 
\nonumber\\ &&
+ 8(10+z) \zeta_2 \ln(z) + 8(5-4 z) \Li_2(1-z)
- 16 (5 +4z) \ln(z) - 160 (1-z)\Biggr\}. 
\end{eqnarray}
In the axial--vector case one has to replace $T_{\rm III, v, int}$ by $T_{\rm III, a, int}$ in 
Eq.~(\ref{eq:PSv})  
\begin{eqnarray}
T_{\rm III, a, int} &=& \Biggl\{
(2+z)\Bigl[
32 S_{1,2}(1-z) - 96 S_{1,2}(-z) - 48 \ln^2(1+z) \ln(z) - 48 \zeta_2 \ln(1+z)
\nonumber\\ &&
+ 40 \ln^2(z) \ln(1+z) - 96 \Li_2(-z)\ln(1+z) \Bigr]
+ 8(1+z)\Bigl[2 \Li_2(-z) + 2 \ln(z) \ln(1+z)
\nonumber\\ &&
+\zeta_2\Bigr]
+8 (2-z) \Li_3(1-z) - 16(6-5z) \Li_3(-z)
- 24 (2-3z)\zeta_3
+ 8 \Li_2(1-z) 
\nonumber\\ &&
+ 8 (2+3z) \Li_2(1-z) \ln(z) 
+ 8(2+5z) \zeta_2 \ln(z) +128 \Li_2(-z) \ln(z) 
-\frac{16}{3} z \ln^3(z) 
\nonumber\\ &&
-4z \ln^2(z) 
-16 \ln(z) - 32(1-z) 
\Biggr\}.
\end{eqnarray}
The result (\ref{eq:PSv}) differs form that in \cite{Berends:1987ab} by the term in \cite{Blumlein:2019srk}, 
Eq.~(8).
Note that the interference term between the diagrams in the upper line and the ones in
the lower line of Figure~\ref{fig:B1} appeared in 
\cite{Berends:1987ab} with the wrong sign, see \cite{Blumlein:2019srk}, Eq.~(9) in the vector case. 
This term is scheme independent and it has been calculated in \cite{SCHELLEKENS} correctly. The axial--vector 
contribution is newly given. This contribution is not contained in Ref.~\cite{Berends:1987ab} but agrees with 
that of Ref.~\cite{Hamberg:1990np}.
\section{\boldmath The interference term between non--singlet and pure singlet corrections}
\label{sec:9}

\vspace*{1mm}
\noindent
The radiator for this process is given by
\begin{eqnarray}
\lefteqn{R_2^{{e^+e^-,\rm NS}-{\rm PS~interf.}}(z,L)} 
 \nonumber\\ &=&
      - \biggl[
            64
            -56 z
            +\frac{16 \big(1+z^2\big)}{1-z}  \text{Li}_2(1-z)
            \nonumber \\ &&
            +\frac{8 \big(5-2 z^2\big)}{1-z} \ln(z)
            +\frac{8 \big(1+z^2\big)}{1-z} \ln^2(z)
      \biggr] L    
      +\frac{8 \big(27-42 z+23 z^2\big)}{1-z}
      \nonumber \\ &&
      -\frac{8 \big(1+2 z^2\big)}{3 (1-z)}  \ln^3(z)
      +\frac{32 \big(1+2 z^2\big)}{1-z} \text{Li}_3(1-z)
      +\frac{64 \big(1+z^2\big)}{1-z} \text{Li}_3(-z)
      \nonumber \\ &&
      +\biggl(
               \frac{32 (1+z)}{1-z}
              -\frac{112 \big(1+z^2\big)}{1-z}  \ln(z)
      \biggr) \zeta_2
      -\frac{80 \big(1+z^2\big)}{1-z} \zeta_3
      -\biggl(
               16 (8-7 z)
               \nonumber \\ &&
              +\frac{16 \big(5-2 z^2\big)}{1-z} \ln(z)
              -\frac{48 \big(1+z^2\big)}{1-z} \ln^2(z)
      \biggr) \ln(1-z)
      \nonumber \\ &&
      +\biggl(
              \frac{8}{(1-z)^2 (1+z)} \big(3+10 z-11 z^2+22 z^3-8 z^4\big)
              +\frac{64 (1+z)}{1-z} \ln(1+z)
      \biggr) \ln(z)
      \nonumber \\ &&
      -\frac{8 (1+z)^2}{1-z} \ln^2(z)
      +\biggl(
               \frac{8 \big(-13+2 z+6 z^2\big)}{1-z}
              -\frac{32 \big(1+z^2\big)}{1-z} \ln(1-z)
              \nonumber \\ &&
              +\frac{16 \big(5+4 z^2\big)}{1-z} \ln(z)
      \biggr) \text{Li}_2(1-z)
      +\biggl(
               \frac{64 (1+z)}{1-z}
              -\frac{32 \big(1+z^2\big)}{1-z} \ln(z)
      \biggr) \text{Li}_2(-z)
      \nonumber \\ &&
      +\frac{128 \big(1+z^2\big)}{1-z} \text{Li}_3(z)
\end{eqnarray}
It is the same in the vector and axial--vector case, since the amplitude squared has one closed fermion line only.
The result differs from that in \cite{Berends:1987ab} by the term in \cite{Blumlein:2019srk}, Eq.~(9).
\section{\boldmath Further terms with no logarithmic enhancement at $O(\alpha^2)$}
\label{sec:10}

\begin{figure}[H]
  \centering
  \includegraphics[width=.24\linewidth,trim={0cm 11.5cm 2.5cm 0}]{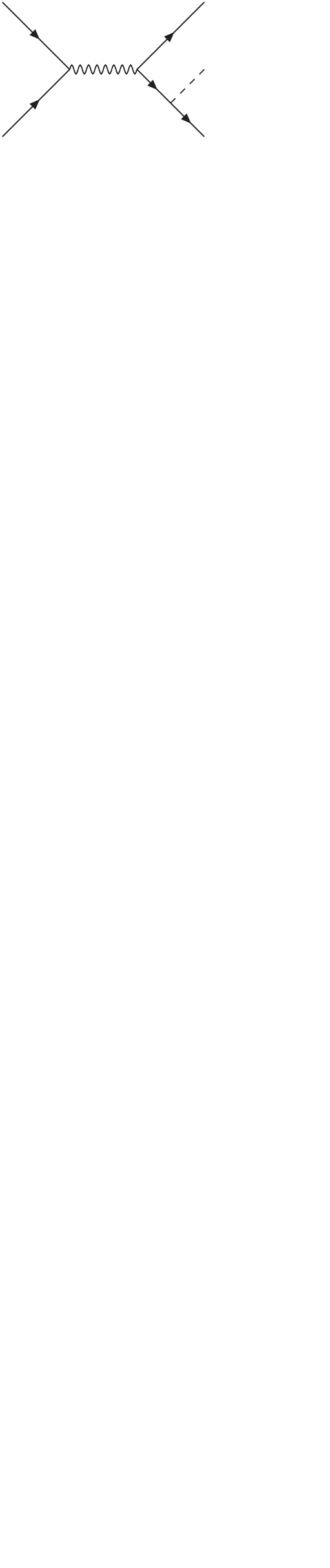}
  \hspace{4cm}
  \includegraphics[width=.24\linewidth,trim={0cm 11.5cm 2.5cm 0}]{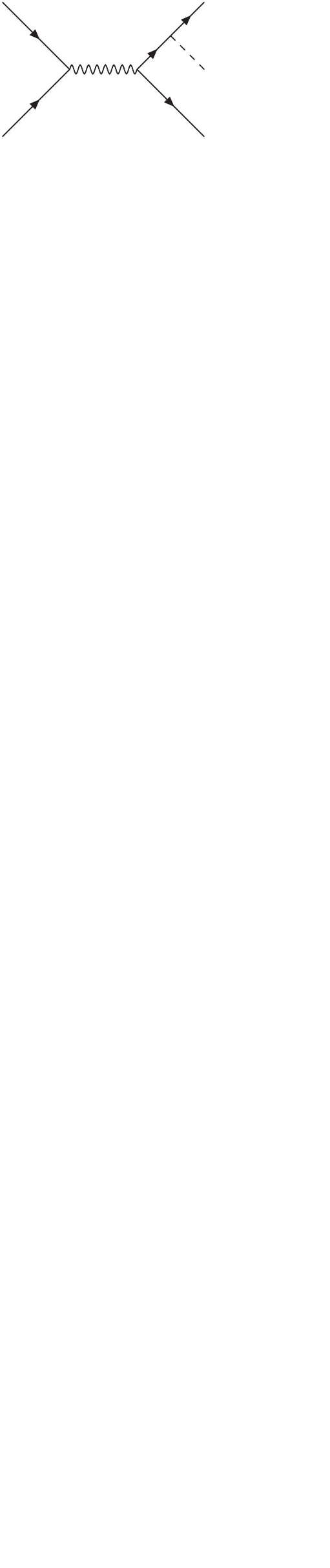}
\vspace*{-17cm}
  \caption{\sf The $e^+e^-$ annihilation graphs into a fermion pair and a gauge boson (process $B$).}
  \label{fig:B}
\end{figure}

\vspace*{1mm}
\noindent
For pure vector couplings there are as well fermion-pair production contributions corresponding to the terms $|B|^2,
BC$ and $BD$ in the case of electrons and for $|B|^2$ for heavier radiated fermions, cf. \cite{Hamberg:1990np}. 
They have
no logarithmic enhancement and were not considered in \cite{Berends:1987ab}. 
The $B$-diagrams are shown in Figure~\ref{fig:B}
The corresponding radiator is given by
\begin{eqnarray}
R _2^{e^+e^- ,\rm v,\rm non log}(z, L) &=& T_{|B|^2} + T_{|BC|+|BD|}
\\
R _2^{e^+e^- ,\rm a,\rm non log}(z, L) &=& T_{|B|^2} + T_{|BC|+|BD|} + T_{|AB|}^A,
\end{eqnarray}
in the vector and axial--vector cases, with
\begin{eqnarray}
T_{|B|^2} &=& 
2 \Bigl\{
-(1+z)^2 
\Bigl[\frac{32}{3} \Li_2(-z) + \frac{16}{3} \zeta_2 - \frac{8}{3} \ln^2(z)
+ \frac{32}{3} \ln(z)\ln(1+z) \Bigr] + \frac{8}{3} (3+3z^2 
\nonumber\\ &&
+4z) \ln(z) + \frac{40}{3}(1-z^2)\Bigr\}
\\
T_{|BC|+|BD|} &=& 
2 \Bigl\{
(1+z^2+3z)\Bigl[32 S_{1,2}(1-z) + 16 \Li_2(1-z) \ln(z) \Bigr]+(1+z)^2 
\Bigl[-48 S_{1,2}(-z) 
\nonumber\\ &&
- 8 \Li_3(-z) +24 \Li_2(-z)
\Bigl[1 + \ln(z) - 2 \ln(1+z)\Bigr] + 12\zeta_2\Bigl[1-2 \ln(1+z) 
\nonumber\\ &&
+ \tfrac{2}{3} \ln(z) \Bigr]
+20 \ln^2(z) \ln(1+z) - 24 \ln^2(1+z) \ln(z) + 24 \ln(z) \ln(1+z)\Biggr] 
\nonumber\\ &&
+ 36(1-z^2) \Li_2(1-z) + \frac{4}{3} (1+z^2+4z) \ln^3(z)
+4(9+11z) \ln(z)+2(6
\nonumber\\ &&
-15z^2-8z) \ln^2(z) +2(27-13z^2-14z)\Bigr\}.
\end{eqnarray}
and
\begin{eqnarray}
T_{|AB|}^A &=& 32 \Biggl\{\frac{1+z^2}{1-z} \ln(z) + 2 z \ln(z) + 3 -z \Biggr\}.
\end{eqnarray}
The contributions of the diagrams $AB$ vanish due to Furry's theorem in the vector case.
We performed the calculation in $D=4$ dimensions keeping the fermion masses, which were set to zero at the 
end of the calculation. We agree with the results of Ref.~\cite{Hamberg:1990np}.
The massive operator matrix element vanishes for these processes and therefore the massive and the massless 
result have to agree according to the factorization theorem postulated in Ref.~\cite{Buza:1995ie}. Due to this,
these contributions have not been included in Ref.~\cite{Blumlein:2011mi}.
\section{Contributions due to Soft Photon Exponentiation beyond \boldmath $O(a^2)$}
\label{sec:11}

\vspace*{1mm}
\noindent
The resummation of the soft corrections has been considered early \cite{Berends:1987ab}. Here the idea is to resum 
first the leading distribution--valued contributions 
\begin{eqnarray}
\frac{d \sigma^{\rm I}}{d s'} = \frac{1}{s} \sigma^{(0)}(s') \left[\delta(1-z) T^{V+S}\left(a,\frac{s}{m^2}\right) 
\exp[\beta 
\ln(\ep)]  + \theta(1-z-\ep) T^H\left(a,\frac{s}{m^2}\right)\right]
\end{eqnarray}
with
\begin{eqnarray}
\beta = 8 a (L - 1)
\end{eqnarray}
and 
\begin{eqnarray}
T^{\rm H}\left(a,\frac{s}{m^2}\right) = \sum_{k=1}^\infty \frac{\beta}{(k-1)!} {\cal D}_{k-1} 
T^{\rm V+S}\left(a,\frac{s}{m^2}\right) + \tilde{T}^{\rm H}\left(a,\frac{s}{m^2}\right).
\end{eqnarray}
The inclusive cross section reads then
\begin{eqnarray}
\label{eq:soft1}
\sigma^{\rm I} = \int_{z_0}^1 \sigma^{(0)}(zs) \left[\beta (1-z)^{\beta-1} T^{\rm V+S}\left(a,\frac{s}{m^2}\right)
+ \tilde{T}^{\rm H}\left(a,\frac{s}{m^2}\right)\right],
\end{eqnarray}
cf.~\cite{Kuraev:1985hb}. The soft--resummed contributions for $O(a^3)$ and higher are
\begin{eqnarray}
\label{eq:soft3}
\sigma^{\rm I, \geq 3, soft} = \int_{z_0}^1 \sigma^{(0)}(zs) \left\{
\beta (1-z)^{\beta-1} - \beta {\cal D}_0(z) - \beta^2 {\cal D}_1(z) \right\} 
T^{\rm 
V+S}\left(a,\frac{s}{m^2}\right).
\end{eqnarray}

One may extend the soft photon exponentiation by including as well the soft production of $e^+e^-$ pairs according
to the non--singlet process described in Section~\ref{sec:5} in the region $z \rightarrow 1$. this modifies
the term $\beta (1-z)^{\beta-1}$ in (\ref{eq:soft1}) to
\begin{eqnarray}
\label{eq:soft2}
\frac{
\exp\left[\tfrac{1}{2}\eta \left(\tfrac{3}{2} - 2 \gamma_E\right)\right]}{\Gamma(\eta)} (1-z)^{\eta-1}
\end{eqnarray}
with
\begin{eqnarray}
\eta = - 6 \ln\left(1 - \frac{4}{3} a L\right),
\end{eqnarray}
cf.~\cite{Gribov:1972ri}. 
The Mellin inversion leading to (\ref{eq:soft2}) has been calculated in 
Ref.~\cite{Gross:1974fm}, see also~\cite{Kuraev:1985hb}.
These are both leading order resummations. One may as well resum the logarithms $\ln^k(z)/z$ in the small $z$
region, cf. e.g.~\cite{Jezabek:1992bx} leading to associated Bessel functions, as has been known in QCD before, 
see e.g.~\cite{RMCK}.

The leading logarithmic orders $O((a  L)^k)$, which are process independent, can be treated rather 
straightforwardly to rather high orders, accounting both for the non--singlet and singlet contributions, cf.~
\cite{Jadach:1988gb,Skrzypek:1992vk,Jezabek:1992bx,
Przybycien:1992qe,Blumlein:1996yz,Arbuzov:1999cq,Arbuzov:1999uq,Blumlein:2004bs,Blumlein:2007kx}. These corrections 
include the 
resummations mentioned 
and do even account for more contributions by resumming as well all collinear contributions according to the 
QED evolution equations. Note, however, that the sub--leading contributions always require the 
inclusion of mass effects due to massive OMEs, cf.~\cite{Blumlein:2011mi}.
\section{Conclusions}
\label{sec:12}

\vspace*{1mm}
\noindent
The $O(a^2)$ initial state radiative corrections to the process $e^+e^- \rightarrow \gamma^*/Z^*$ have been
computed in a direct calculation without neglecting the electron mass against the $s$--channel energy of the process.
The expansion in the ratio $m_e^2/s$ has been only performed in a very late stage of the calculation by controlling the
result based on precision numerics in {\tt mathematica} with the complete result. The corrections can be grouped
into four main processes, I--IV, as already done in Ref.~\cite{Berends:1987ab}, with the addition of non--logarithmic 
terms 
and terms due to soft--exponentiation for the contributions beyond $O(a^2)$. Furthermore, one has to account for 
differing axial--vector contributions in some of the channels. For the  processes I--IV 
we find differing results for the non--logarithmic terms of $O(a^2)$ given in \cite{Berends:1987ab}, while 
we agree in the logarithmic 
contributions and those of $O(a)$. On the other hand, we agree with the results of Ref.~\cite{Blumlein:2011mi}. In 
the case of 
process II
we agree with Refs.~\cite{Berends:1987ab,Kniehl:1988id} if the initial state fermion radiation concerns 
$\mu^+\mu^-$ or heavier lepton or quark pairs. We also agree for the pure--singlet interference terms with a 
result in Ref.~\cite{SCHELLEKENS}. Furthermore, we agree with non--logarithmic corrections derived first for the 
Drell--Yan process \cite{Hamberg:1990np}.

The present calculation proofs, here for QED, that the massive Drell--Yan process factorizes, and we revise 
an earlier doubt in Ref.~\cite{Blumlein:2011mi}. The present rather voluminous calculation has been the only 
way to establish this. Fortunately, mathematical methods are now available to perform the corresponding 
integrals analytically and allow to represent them as iterated integrals of square root--valued alphabets, 
carrying real parameters. It is this representation which finally allows the controlled limit $m_e^2/s 
\rightarrow 0$ for the power corrections. A part of the integrals are incomplete elliptic integrals and 
generalizations thereof, which does not lead to a further sophistication, since the corresponding integrals 
are still iterative. Numerical illustrations of the present results have been given in Ref.~\cite{Blumlein:2019pqb}.
already.

The numerical accuracy to which both the $Z$ boson mass and width are planned to be measured at the 
FCC\_ee is rather high. It amounts to $\sim 100$~keV systematic uncertainty, with a much higher statistical 
precision. It is clear from Ref.~\cite{Blumlein:2019pqb} that the $O(a^2)$ ISR corrections will not yet be 
sufficient to cope with this accuracy. Therefore, even higher order corrections have to be calculated to
sub--leading levels, cf.~\cite{BDS}.
\appendix
\section{The phase space integrals}
\label{sec:A}

\vspace*{1mm}
\noindent
In the following we describe the parameterization of the phase space integrals both for 
the case of fermion pair and photon pair radiation, followed by explicit expressions 
obtained after the angular integrations. Here the setup is similar to that used in 
\cite{Berends:1987ab,SCHELLEKENS}.
\subsection{Fermion Pair Radiation} 
\label{sec:A1}
For massive fermion pair radiation we only encounter $2 \to 3$ scattering with the kinematics
\begin{align}
	p_- + p_+ &= q + k_- + k_+
\end{align}
with
\begin{align}
	(p_- + p_+)^2 &= s ,
\nonumber \\
	q^2 &= \sp ,
\nonumber \\
	p_-^2 = p_+^2 &= k_-^2 = k_+^2 = m^2 .
\end{align}
We also introduce the invariants
\begin{align}
	( k_+ + q )^2 &= s_3 ,
\\
	( k_- + q )^2 &= s_4 ,
\\
	( k_- + k_2 )^2 &= \spp,
\end{align}
which satisfy the identity 
\begin{align}
	s_3 + s_4 + \spp &= s + \sp + m^2  .
\end{align}
The phase space integral is given by
\begin{align}
	\int d \text{PS}_3 &= \frac{1}{(2\pi)^6} \int d^4 q \int d^4 k_- \int d^4 k_+
	\biggl\{
		\delta \left( q^2 - \sp \right)
		\delta \left( k_-^2 - m^2 \right)
\nonumber \\ & \times
		\delta \left( k_+^2 - m^2 \right)
		\delta^{(4)} \left( p_- + p_+ - q - k_- - k_+ \right)
	\biggr\}
\nonumber \\ 
	&= \frac{1}{(2\pi)^6} \int d^4 k_1 \int d^4 k_2 
	\biggl\{ 
		\delta( \left[ p_- + p_+ - k_1 - k_2 \right]^2 - \sp) 
		\delta(k_1^2 - m^2) 
\nonumber \\ & \times
		\delta(k_2^2 - m^2) 
	\biggr\} 
\nonumber \\
	&= \frac{1}{(2\pi)^5} 
	\int d k_1^0 
	\int d |\vec{k}_1| 
	\int d \cos(\chi) 
	\int d k_2^0 \int d |\vec{k}_2| 
	\int\limits_{-1}^{1} d \cos(\theta) 
	\int\limits_{0}^{2\pi} d\phi 
\nonumber \\ & \times 
	\biggl\{ 
		|\vec{k}_1|^2 |\vec{k}_2|^2 
		\frac{\delta( \cos(\chi) - \cos(\chi_0) )}{2 |\vec{k}_1| |\vec{k}_2|} 
		\frac{\delta( |\vec{k}_1| - \sqrt{ (k_1^0)^2 - m^2 })}{ 2 |\vec{k}_1| }
		\frac{\delta( |\vec{k}_2| - \sqrt{ (k_2^0)^2 - m^2 })}{ 2 |\vec{k}_2| } 
	\biggr\} 
\nonumber \\ 
	&= \frac{1}{4 (2\pi)^5} 
	\int d k_1^0 
	\int d k_2^0 
	\int\limits_{-1}^{1} d \cos(\theta) 
	\int\limits_{0}^{\pi} d\phi 
\nonumber \\
	&= \frac{1}{(4\pi)^4} \frac{1}{2 \pi s} 
	\int d s_3 
	\int d s_4 
	\int\limits_{-1}^{1} d \cos(\theta) 
	\int\limits_{0}^{\pi} d\phi 
\nonumber \\
	&= \frac{1}{(4\pi)^4} \frac{1}{2 \pi s} 
	\int d \spp 
	\int d s_3 
	\int\limits_{-1}^{1} d \cos(\theta) 
	\int\limits_{0}^{\pi} d\phi. 
\end{align}
In deriving these relations, the identities
\begin{align}
\delta(\left[ p_- + p_+ - k_1 - k_2 \right]^2 - \sp) 
	&=\delta( s - \sp - 2 \sqrt{s} ( k_1^0 + k_2^0 ) + 2 m^2 + 2 k_1.k_2 ) 
\nonumber \\
	&= \delta( s - \sp - 2 \sqrt{s} ( k_1^0 + k_2^0 ) + 2 m^2 + 2 k_1^0 k_2^0 - 2 |\vec{k}_1| |\vec{k}_2| \cos(\chi) ) 
\nonumber \\
	&= \frac{1}{2 |\vec{k}_1| |\vec{k}_2| } \delta( \cos(\chi) - \cos(\chi_0) ) ,
\end{align}
with
\begin{align}
\cos(\chi_0) &= \frac{s - \sp + 2 m^2 - 2 \sqrt{s} ( k_1^0 + k_2^0 ) + 2 k_1^0 k_2^0}{2 |\vec{k}_1| |\vec{k}_2|} ,
\end{align}
were used.
The integration variables are transformed according to
\begin{align}
	s_3 &= ( k_2 + q )^2 = ( p_- + p_+ - k_1 )^2 = s + m^2 - 2 \sqrt{s} k_1^0 ,
\nonumber \\
	s_4 &= ( k_1 + q )^2 = ( p_- + p_+ - k_2 )^2 = s + m^2 - 2 \sqrt{s} k_2^0 ,
\nonumber \\
	d s_3 &= - 2 \sqrt{s} d k_1^0 ,
\nonumber \\
	d s_4 &= - 2 \sqrt{s} d k_2^0 ,
\end{align}
and the symmetry of the angular integration allows to transform
\begin{align}
	\int\limits_{-1}^{1} d \cos(\theta) \int\limits_{0}^{2\pi} d\phi 
	&= 2 \int\limits_{-1}^{1} d \cos(\theta) \int\limits_{0}^{\pi} d\phi .
\end{align}
The phase space boundaries are given by
\begin{eqnarray}
4 m^2 <& \spp &< ( \sqrt{s} - \sqrt{\sp} )^2 ,
\\
s_3^- <& s_3 &< s_3^+ ,
\end{eqnarray}
where the explicit expressions for $s_3^-$ and $s_3^+$ are given by
\begin{align}
s_3^\pm &= \frac{1}{2} \left( s + \sp - \spp + 2 m^2 \pm \sqrt{ 1 - \frac{4m^2}{\spp} } \lambda^{1/2} (s,\sp,\spp) \right).
\end{align}
We can also change the order of integration in which case we obtain
\begin{eqnarray}
( \sqrt{s} - m )^2 <& s_3  &< ( \sqrt{\sp} - m )^2 ,\\
\left. \spp \right.^- <& \spp &< \left. \spp \right.^+ 
\end{eqnarray}
with the explicit expressions
\begin{align}
\left. \spp \right.^{\pm} &= \frac{1}{2 s_3} \left( (s-s_3)(s_3-\sp) + m^2(s+2s_3+\sp) - m^4 \pm \lambda^{1/2}(s,s_3,m^2) \lambda^{1/2}(\sp,s_3,m^2) \right) .
\end{align}
We can use the following parameterization of the vectors:
\begin{eqnarray}
	p_- &=& \frac{\sqrt{s}}{2} \begin{pmatrix} 1, & 0,& 0,& \beta \end{pmatrix} 
\nonumber \\
	p_+ &=& \frac{\sqrt{s}}{2} \begin{pmatrix} 1, & 0,& 0,& - \beta \end{pmatrix} 
\nonumber \\
	k_1 &=& \begin{pmatrix} k_1^0 , & 0 , & |\vec{k}_1| s(\theta) , & |\vec{k}_1| c(\theta) \end{pmatrix}  \\
	k_2 &=& \begin{pmatrix} k_2^0 , & |\vec{k}_2| s(\phi)s(\chi_0) , & |\vec{k}_2| \left( c(\chi_0)s(\theta) - c(\theta)c(\phi)s(\chi_0) \right) , & |\vec{k}_2| \left( c(\theta) c(\chi_0) + c(\phi) s(\theta) s(\chi_0) \right) \end{pmatrix} 
\nonumber \\
	q &=& p_- + p_+ - k_1 - k_2
\end{eqnarray}
with the abbreviation $c(x)= \cos(x)$ and $s(x)=\sin(x)$.
The missing components of the vectors are given by
\begin{align}
	k_1^0 &= \frac{s - s_3 + m^2}{2s}, &
	|\vec{k}_1| &= \frac{\lambda^{1/2}(s,s_3,m^2)}{2s},  
\nonumber \\
	k_2^0 &= \frac{s - s_4 + m^2}{2s}, &
	|\vec{k}_2| &= \frac{\lambda^{1/2}(s,s_4,m^2) }{2s}. 
\end{align}
The direction of the 3-vector component of $k_2$ is achieved by rotating $\vec{k}_1$ with 
angle $\chi_0$ around the $x$-axis and then with angle $\phi_0$ around $k_1$.
It is convenient to transform to the dimensionless variables
\begin{eqnarray}
t = \frac{\sp}{s}, \quad
x = \frac{s_3}{s}, \quad
y = \frac{4 m^2}{\spp},
\end{eqnarray} 
in the explicit calculations.
Since all involved particles are massive, the phase space integrals are convergent and do not need any kind of 
regularization.
\subsection{Photon Radiation} 
\label{sec:A2}
The $2 \to 3$ scattering can be very similarly parameterized as before.
However, the replacements $k_- \to k_1$ and $k_+ \to k_2$ with
\begin{align}
	k_1^2 &= k_2^2 = 0 
\end{align}
have to be made.
Therefore, the limit $m \to 0$ has to be taken in the expressions given in the previous section.
We will give the explicit expressions for completeness.

The phase space integral reads
\begin{align}
\int d \text{PS}_3 &= \frac{1}{(2\pi)^6} \int d^4 q \int d^4 k_- \int d^4 k_+
	\biggl\{
		\delta \left( q^2 - \sp \right)
		\delta \left( k_-^2 - m^2 \right)
\nonumber \\ & \times
		\delta \left( k_+^2 - m^2 \right)
		\delta^{(4)} \left( p_- + p_+ - q - k_- - k_+ \right)
	\biggr\}
\nonumber \\ 
	&= \frac{1}{(4\pi)^4} \frac{1}{2 \pi s} 
	\int d s_3 
	\int d s_4 
	\int\limits_{-1}^{1} d \cos(\theta) 
	\int\limits_{0}^{\pi} d\phi 
\end{align}
with the explicit parameterization of the vectors
\begin{eqnarray}
	p_- &=& \frac{\sqrt{s}}{2} \begin{pmatrix} 1, & 0,& 0,& \beta \end{pmatrix} 
\nonumber \\
	p_+ &=& \frac{\sqrt{s}}{2} \begin{pmatrix} 1, & 0,& 0,& - \beta \end{pmatrix} 
\nonumber \\
	k_1 &=& \frac{s-s_3}{2\sqrt{s}} \begin{pmatrix} 1 , & 0 , &  s(\theta) , & c(\theta) \end{pmatrix}  \\
	k_2 &=& \frac{s-s_4}{2\sqrt{s}} \begin{pmatrix} 1 , & s(\phi)s(\chi_0) , & \left( c(\chi_0)s(\theta) - c(\theta)c(\phi)s(\chi_0) \right) , & \left( c(\theta) c(\chi_0) + c(\phi) s(\theta) s(\chi_0) \right) \end{pmatrix} 
\nonumber \\
	q &=& p_- + p_+ - k_1 - k_2 .
\end{eqnarray}
The angle between the two photons is given by
\begin{align}
	\cos (\chi_0) &= 1 - \frac{2 s \spp}{(s-s_3)(s-s_4)} .
\end{align}
The phase space boundaries simplify to
\begin{eqnarray}
 	\frac{s \sp}{s_3} \leq & s_4 & \leq s + \sp - s_3 ,
\\ \nonumber
	\sp \leq & s_3 & \leq s .
\end{eqnarray}
They are symmetric in $s_3$ and $s_4$.

It is also possible to only radiate one additional photon.
In this case the phase space for $2 \to 2$ scattering is needed.
Using the kinematics
\begin{align}
	p_- + p_- &= q + k
\end{align}
with $k^2 = 0$,
it is given by
\begin{align}
	\int d \text{PS}_2 &= 
	\int d^4 q \int d^4 k \delta(s-\sp) \delta( k^2 ) \delta^{(4)}(p_- + p_+ - q - k)  
\\ \nonumber 
	&= \frac{1}{(4\pi)^2} \frac{2}{s-\sp} \int\limits_{-1}^{1} d \cos(\theta) .
\end{align}
In this case the vectors can be parameterized by
\begin{eqnarray}
	p_- &=& \frac{\sqrt{s}}{2} \begin{pmatrix} 1, & 0, & 0, & \beta \end{pmatrix} , 
\\ \nonumber
	p_+ &=& \frac{\sqrt{s}}{2} \begin{pmatrix} 1, & 0, & 0, & - \beta \end{pmatrix} ,
\\ \nonumber 
	k   &=& \frac{s-\sp}{2\sqrt{s}} \begin{pmatrix} 1, & 0, & \sin(\theta), & \cos(\theta) \end{pmatrix} ,
\\ \nonumber
	q &=& p_- + p_+ - k .
\end{eqnarray}

\subsection{The Angular Integrals}
\label{sec:A3}

For the photon emission graphs we find the following denominators
\begin{align}
	D_1 &= (p_- - k_2)^2 - m^2
,
&
	D_2 &= (p_- - k_1)^2 - m^2
,
\nonumber \\
	D_3 &= (q - p_+)^2 - m^2
,
&
	D_4 &= (q - p_-)^2 - m^2
,
\nonumber \\
	D_5 &= (p_+ - k_2)^2 - m^2
,
&
	D_6 &= (p_+ - k_1)^2 - m^2
.
\end{align}

For the angular integrals we again want to map to the angular integrals of the form
\begin{align}
	I^{d=4}_{l,k} &= \int\limits_{0}^\pi d\theta \int\limits_{0}^{\pi} d\phi \frac{ \sin(\theta) }
{\left[ a + b \cos(\theta) 
\right]^l } \frac{1}{ \left[A + B \cos(\theta) + C \sin(\theta) \cos(\phi) \right]^k }  
\label{eq:angintsd4}
\end{align}
For some denominator structures we have to use partial fractioning.
Some cases are trivial, like
\begin{align}
	\frac{1}{D_2 D_6} &= \frac{1}{s_3 - s} \left( \frac{1}{D_1} + \frac{1}{D_5} \right) ,
\\
	\frac{1}{D_3 D_4} &= \frac{1}{\spp - s} \left( \frac{1}{D_3} + \frac{1}{D_4} \right) ,
\\
	\frac{1}{D_1 D_5} &= \frac{1}{s_4 - s} \left( \frac{1}{D_5} + \frac{1}{D_6} \right) .
\end{align}
The more involved ones read
\begin{align}
	\frac{1}{D_1 D_2 D_3} &= \frac{1}{\spp} \left( \frac{1}{D_1 D_2} - \frac{1}{D_1 D_3} - \frac{1}{D_2 D_3} \right),
\\
	\frac{1}{D_2 D_3 D_5} &= \frac{1}{\sp-s_3} \left( \frac{1}{D_2 D_3} + \frac{1}{D_2 D_5} - \frac{1}{D_3 D_5} \right),
\\
	\frac{1}{D_1 D_3 D_6} &= \frac{1}{\sp-s_4} \left( \frac{1}{D_1 D_6} + \frac{1}{D_1 D_3} - \frac{1}{D_3 D_6} \right),
\\
	\frac{1}{D_1 D_4 D_6} &= -\frac{1}{\sp-s_3} \left( \frac{1}{D_1 D_6} - \frac{1}{D_1 D_4} + \frac{1}{D_4 D_6} \right),
\\
	\frac{1}{D_2 D_4 D_5} &= \frac{1}{\sp-s_4} \left( \frac{1}{D_2 D_5} + \frac{1}{D_4 D_5} - \frac{1}{D_2 D_4} \right)
\\
	\frac{1}{D_4 D_5 D_6} &= \frac{1}{\spp} \left( \frac{1}{D_5 D_6} - \frac{1}{D_4 D_5} - \frac{1}{D_4 D_5} \right).
\end{align}
For some combinations of denominators we have to interchange the parameterizations of $k_-$ and $k_+$
in order to arrive at angular integrals of the form \eqref{eq:angintsd4}.

If either $l$ or $k$ are negative we can use the relations given in 
Eqs.~(\ref{eq:massiveangularint},\ref{eq:massiveangularint2}) for $D=4$ to arrive at the angular integrals.
If both indices are negative we were not able to find a closed form in $D$ dimensions.
For $D=4$ we find
\begin{align}
	I^{d=4}_{-2,-2} &= 2 \pi \frac{b^4 A^4 -2 a b^3 A^3 B - 2 a b A B (a^2 - 2 b^2)(B^2+C^2) - b^2 A^2 \bigl( 2 b^2 B^2 - a^2 ( 2 B^2 - C^2) \bigr)}{(a^2-b^2)(A^2-B^2-C^2)X^2} 
\nonumber \\ &
- \frac{(B^2 + C^2)\bigl(2 a^2 b^2 B^2 + b^4 C^2 - a^4 (B^2+C^2) \bigr)}{(a^2-b^2)(A^2-B^2-C^2)X^2}
\nonumber \\ &
- b \pi \frac{2 b^2 A^2 B + b^2 B C^2 + 2 a^2 B ( B^2 + C^2 ) - a b A ( 4B^2 +3 C^2)}{X^{5/2}} \ln\left( \frac{a A - b B + \sqrt{X}}{a A - b B - \sqrt{X}} \right)
,
\\
	I^{d=4}_{-2,-1} &= \frac{2b (b A - a B ) \pi}{(a^2-b^2) X} + \pi \frac{a(B^2+C^2)-b A B}{X^{3/2}} \ln\left( \frac{a A - b B + \sqrt{X}}{a A - b B - \sqrt{X}} \right)
,
\\
	I^{d=4}_{-1,-2} &= \frac{2 \pi ( a ( B^2+C^2 ) - b A B )}{(A^2-B^2-C^2)X} + \frac{b (b A - a B)\pi}{X^{3/2}} \ln\left( \frac{a A - b B + \sqrt{X}}{a A - b B - \sqrt{X}} \right)
,
\\
	I^{d=4}_{-1,-1} &= \frac{\pi}{\sqrt{X}} \ln\left( \frac{a A - b B + \sqrt{X}}{a A - b B - \sqrt{X}} \right)
,
\end{align}
with $X= (a A - b B)^2 - (a^2 - b^2)(A^2 - B^2 - C^2)$.
Note that we agree with the results given in \cite{Beenakker:1988bq,Hamberg:1990np,Bojak:2000eu}.

\vspace*{5mm}
\noindent
{\bf Acknowledgments}

\noindent
This paper is dedicated to the memory of our colleague W.L.~van Neerven. We would like to thank J.C.~Collins, B.~Kniehl,
J.H.~K\"uhn, A.~Maier, P.~Marquard, G.~Passarino, C.~Schneider and G.~Sterman for discussions. This project has received
funding from the European Union's Horizon 2020 research and innovation programme under the Marie
Sk\l{}odowska-Curie grant
agreement No. 764850, SAGEX, and COST action CA16201: Unraveling new physics at the LHC through the precision frontier
and from the Austrian FWF grants P 27229 and P 31952 in part. The diagrams have been drawn using {\tt Axodraw}
\cite{Vermaseren:1994je}.

\end{document}